\documentclass[12pt]{article}
\usepackage[margin=1in]{geometry}
\usepackage[table]{xcolor}
\usepackage{graphicx}
\usepackage{lineno,scalerel,mathrsfs,amsmath,amssymb,xcolor,chemarr,titlesec,subfigure,amsthm,zref-xr,authblk,setspace,epstopdf}
\usepackage[euler]{textgreek}
\newcommand{\lra}{\mathop{\longrightarrow}\limits}
\newcommand\hill{\mathcal{H}}
\newcommand\mr{\mathrm}
\renewcommand{\textbf}[1]{{\sffamily\bfseries #1}}
\newtheorem{theorem}{\textbf{Theorem}}
\newcommand\s[1]{^{\scaleto{({#1})}{6pt}}}
\newcommand\su[1]{_{\scaleto{{#1}}{4pt}}}

\renewcommand{\figurename}{{\bfseries Figure}}
\renewcommand{\tablename}{{\bfseries  Table}}

\usepackage[compress,numbers]{natbib}
\usepackage[font=small,labelfont=sf,bf]{caption}
\titleformat{\section}{\Large\bfseries\sffamily}{S\thesection}{1em}{}
\titleformat{\subsection}{\large\bfseries\sffamily}{S\thesubsection}{1em}{}
\titleformat{\subsubsection}{\bfseries\sffamily}{S\thesubsubsection}{1em}{}
\renewcommand{\paragraph}[1]{\vspace{1em}\noindent{\bfseries\sffamily #1}.}

\title{\bfseries \sffamily Epigenetic factor competition reshapes the  EMT landscape}

\author[a,b]{M. Ali Al-Radhawi} 
 
\author[b,c,f]{Shubham Tripathi}
\author[d,h]{Yun Zhang}
\author[a,e,g,*]{Eduardo D. Sontag}
\author[b,e,f,*]{Herbert Levine}

\affil[a]{\footnotesize Department of Electrical and Computer Engineering, Northeastern University, Boston, MA 02115.}
\affil[b]{\footnotesize Center for Theoretical Biological Physics, Northeastern University, Boston, MA 02115.}
\affil[c]{\footnotesize PhD Program in Systems, Synthetic, and Physical Biology, Rice University, Houston, TX 77005.}
\affil[d]{\footnotesize Whitehead Institute for Biomedical Research, Cambridge, MA 02142, USA.}
\affil[e]{\footnotesize Department of Bioengineering, Northeastern University, Boston, MA 02115.}
\affil[f]{\footnotesize Department of Physics, Northeastern University, Boston, MA 02115.}
\affil[g]{\footnotesize Laboratory of Systems Pharmacology, Harvard Medical School, Boston, MA 02115.}
\affil[h]{\footnotesize State Key Laboratory of Molecular Oncology, National Cancer Center/National Clinical Research Center for Cancer/Cancer Hospital, Chinese Academy of Medical Sciences and Peking Union Medical College, Beijing 100021, China.}
\date{August 2nd, 2022}
\begin{document}
	\maketitle 
	\begin{abstract}{ The emergence of and transitions between distinct phenotypes in isogenic cells can be attributed to the intricate interplay of epigenetic marks, external signals, and gene regulatory elements. These elements include chromatin remodelers, histone modifiers, transcription factors, and regulatory RNAs. Mathematical models known as Gene Regulatory Networks (GRNs) are an increasingly important tool to unravel the workings of such complex networks. In such models, epigenetic factors are usually proposed to act on the chromatin regions directly involved in the expression of relevant genes. However, it has been well-established that these factors operate globally and compete with each other for targets genome-wide. Therefore, a perturbation of the activity of a regulator can redistribute epigenetic marks across the genome and modulate the levels of competing regulators. In this paper, we propose a conceptual and mathematical modeling framework that incorporates both local and global competition effects between antagonistic epigenetic regulators in addition to local transcription factors, and show the counter-intuitive consequences of such interactions. We apply our approach to recent experimental findings on the Epithelial-Mesenchymal Transition (EMT). We show that it can explain the puzzling experimental data as well provide new verifiable predictions.}
	\end{abstract}

	 {M}ulticellular organisms start from the mitotic division of a single cell, and then proliferate and  \emph{differentiate} into increasingly more specialized \emph{lineages}. Within the central dogma of molecular biology,   differences between cells in different lineages can be explained by different patterns of gene activity \cite{ama_book}. Even within the same lineage, different genes can be activated depending on the external signals, environmental factors, or internal stochasticity. Hence, precise mechanisms for gene regulation must exist within the cell. Such mechanisms are commonly ascribed to a multitude of processes involving epigenetic, transcriptional, translational and post-translational regulatory elements. However, the manner in which the different layers of regulation interact is far from being fully understood. 
		
		The complex interplay between transcriptional and epigenetic control is strikingly evident in the regulation of the epithelial-mesenchymal transition (EMT), a process that allows cells to lose cell-cell adhesion and apico-basal polarity, and acquire migratory and invasive traits \cite{emt_2016}. EMT is essential for embryonic development and wound healing. However, dysregulated EMT is a key contributor to cancer mortality, playing crucial roles in metastasis and emergence of drug resistance. Epigenetic processes have been known to act both upstream and downstream of the core EMT gene regulatory circuit \cite{emt_2016} and various epigenetic modifications are known to alter the expression of transcription factors and micro-RNAs involved in EMT control \cite{Wu2012, scheel, chaffer}. In turn, the global epigenetic state of cells can be altered by the induction of one or more EMT transcription factors \cite{scheel}. 
		
		The complexity of the epigenetic-transcriptional interplay in EMT has been highlighted in a recent study by Zhang \emph{et al}.\ \cite{yun22}. This work have shown that the knockout of different histone methyltransferases in the human mammary epithelial cell line HMLER could induce two distinct trajectories of EMT, characterized by distinct and unexpected changes in gene expression profiles. It would clearly be useful to interpret these results within the context of increasing powerful mathematical models of gene regulation and the EMT process. 

		{A popular framework to model and predict the consequences of interactions between various biomolecular regulators uses the concept of} a gene regulatory network (GRN) \cite{karlebach08}.  {Such descriptions can recapitulate the phenotypic heterogeneity that may be exhibited by genotypically identical cells by taking into account the differences in the configuration of the network, initial conditions, and external factors \cite{huang05, racipe}.} Early GRN models mainly considered the interactions between transcription factors (TFs) and promoters \cite{kauffman71,alon,karlebach08}. GRNs were later expanded to include the activities of non-coding RNAs such as micro-RNAs \cite{lu13b}, DNA methylation \cite{olariu16,chen21}, and histone modifiers \cite{thalheim17,Zhang19,Alarcon21,bruno22}. Complementary mathematical models have focused on the effect of antagonistic epigenetic factors (EFs) acting on the histone tails \cite{sneppen2019,zhao21}{, or the interaction between repressive histone modifiers and transcription factors \cite{ringrose20}}.
		
		In the context of modeling EMT, most of the systems biology modeling effort  has focused exclusively on the transcriptional and translational dynamics \cite{systemsbiologyEMT}. Conversely,  mathematical modeling of the underlying epigenetic processes has largely been limited to coarse-grained phenomenological approaches \cite{Jia2019, Jia2020}. Thus, recent high-resolution characterizations of the epigenetic and transcriptional changes during EMT and their response to epigenetic and transcriptional perturbations \cite{karacosta19,scheel, yun22} provide both a need and an opportunity for the development of new models that can shed light on the principles governing the complex transcriptional-epigenetic interplay.

		In summary, the aforementioned formulations, both for general GRNs and for EMT, do not consider genome-wide effects. Instead, they focus on the genomically \emph{local} interactions of the regulatory factors with a single gene or a small set of genes. Although such an assumption might be justified in many cases, this ignores the fact that many regulatory factors, and especially epigenetic ones, act globally and can have hundreds or thousands of targets. Furthermore, such factors compete with each other, and perturbations to the activity or expression level of one of them can have considerable off-target effects, as will be reviewed next.
		
		\subsection*{The Polycomb and Trithorax groups of epigenetic factors}
		In this paper, we   focus on the well-documented antagonism between the Polycomb (PcG) and Trithorax (TrX) protein groups. These protein families modulate histone tails that help maintain genes in  silenced  or active states, and that act globally to regulate numerous cellular processes \cite{Piunti16, schuettengruber17}. PcG and TrX act antagonistically, where the first is usually associated with silencing, while the latter is associated with activation \cite{Piunti16, schuettengruber17}. For instance, Polycomb Repressive Complex 2  (PRC2), a PcG protein, is responsible for trimethylating Histone H3 lysine 27 (H3K27) to mark genes for silencing \cite{Margueron11, conway15}. PRC2 has been reported to have more than 1000 targets in a single human embryonic fibroblast cell \cite{bracken06} and it is estimated  to target at least 10\% of the genes in Embryonic Stem Cells (ESCs) \cite{mohn08}.
		
		On the other hand, the COMPASS family of proteins (a subfamily of TrX proteins) is involved in depositing activating  methylation marks. In particular, KMT2A/B (MLL1/2), KMT2C/D (MLL3/4), and SETD1A/B (KMT2F/G) deposit methylation marks at Histone H3 lysine 4 (H3K4) \cite{Shilatifard12}. These proteins differ in the genomic region targeted: SETD1A/B trimethylate H3K4 around transcription start sites (TSSes) \cite{wu08}, KMT2C/D mono-methylate H3K4 around enhancer elements \cite{hu13,froimchuk17}, and KMT2A/B deposit H3K4me2 and H3K4me3 marks at developmental genes \cite{hu17}. It should be noted that these factors often have overlapping effects depending on the context. For instance, knockout of KMT2D causes genome-wide disappearance of H3K4me3 in mice B cells \cite{zhang15} and brain-cells \cite{dhar18}, and it has been shown to be essential for the maintenance of H3K4me2 marks in mice cardiomyocytes \cite{ang16}. In addition, KMT2D is required for acetylating H3K27 in conjunction with CREBBP and EP300 \cite{wang16}, where H3K27ac is an activating mark that is mutually exclusive with the silencing mark H3K27me3. Similar to PRC2, TrX group proteins act genome wide. For instance, the transcription of 1,200 genes has been shown to be dependent on KMT2B in mouse embryonic stem cells \cite{Douillet20}, while KMT2D was shown to bind to 4,880 genes in mice cardiomyocytes \cite{ang16}. In summary, PcG and TrX proteins \textit{act genome wide}, \textit{deposit marks on similar histone sites}, and \textit{have opposing functions}.
		
		\subsection*{Epigenetic factor competition}
		As reviewed above, antagonistic EFs deposit functionally opposing histone marks. Do they compete for the same (or nearby) genomic sites? PcG and TrX proteins are recruited to genes by regulatory sequences known as polycomb response elements (PREs) and trithorax response elements (TREs), respectively \cite{steffen14}. Existing evidence in Drosophila  shows that PREs are also TREs and that PcG/TrX proteins compete for them \cite{steffen14,sneppen2019}. In addition, activating methylation marks (e.g, H3K27ac, H3K4me3 and H3K36me3) inhibit PRC2's ability to methylate H3K27 \cite{klymenko04,tie09,schmitges11}. On other hand, PRC2's activity reduces the ability of CREBBP/EP300/KMT2D to deposit H3K27ac  activating marks \cite{pasini10, banerjee16}. While it might be possible for H3K4me3 and H3K27me3 to exist in the same vicinity (a phenomenon known as bivalency \cite{blanco20}), they are mutually exclusive on the same histone tail \cite{shema16}.  {Alternative mechanisms of competition include PRC2 acting indirectly on nearby nucleosomes by recruiting other factors to remove activating marks \cite{pasini08,jin17} .} Therefore, it is usually assumed that PcG and TrX counteract each other genome-wide \cite{schuettengruber17, sneppen2019}. 
		
		{In addition to  direct competition with other EFs, PRC2 is antagonistic to active transcription.  PRC2 activity leads to chromatin compaction \cite{guo21},  which makes it harder for activating TFs to access their target sites. On the other hand, PRC2 can read the epigenetic context to avoid acting on genes that are transcriptionally active \cite{Holoch17,wiles17,davidovich21}. Possible mechanisms include PRC2 binding to nascent RNA \cite{beltran16}, relative aversion to open chromatin \cite{yuan12}, likely competition with transcription factors \cite{davidovich21}, enhancer-PRE communication \cite{perez11}, among others \cite{ringrose20}.}
		
		\subsection*{PcG dilution and redistribution upon suppression of competitors}
		Since PcG and its competitors vie for similar genomic sites, knockout of one factor can have far-reaching off-target effects via redistribution or dilution of its competing factors. Therefore, new genes might get activated or silenced. Such knockout experiments have been conducted in the literature with a particular attention to PcG proteins and the corresponding histone mark (H3K27me3). Below, we review experiments that provide evidence supporting the sequestration and redistribution of PcG proteins.
		
		The protein MES4 is an H3K36 methyltransferase that is antagonistic to PcG. In \cite{gaydos12}, loss of MES4  in \textit{C. elegans} caused a  reduction of H3K27me3 levels (deposited by PcG) at its target sites. Meanwhile, genomic sites that lost the antagonistic mark H3K36me3 gained H3K27me3. Similarly in mice \cite{lu16}, an H3K36M mutant inhibits the activity of H3K36 methyltransferases. The authors observed sequestration of PRC2 as evidenced by increased levels of chromatin-bound EZH2 and SUZ12 (sub-components of PRC2). In addition, many genes lost H3K27me3 and their expression levels increased. This indicated that the loss of H3K36 methylation can provide new substrates for PRC2. Redistribution of H3K27me3 was also observed in H3K36M mutants in \textit{Drosophila} \cite{chaouch21}. In a recent investigation, rapid depletion of the BAF complex (a chromatin remodeler that is antagonistic to PcG) redistributed PcG from highly occupied domains to new genomic sites in mouse ESCs \cite{weber21}. Additional pieces of evidence are included in the Discussion section.
		
		Overall, PcG proteins are highly sensitive to perturbations to other EFs and epigenetic marks. Such interventions reshape the global epigenetic landscape leading to aberrant changes in transcription. This, in turn, has been identified as a contributing factor to various malignancies \cite{mortimer19, khazaei20}.

		\subsubsection*{\upshape Organization of the paper} %
		We first propose a general modeling framework, and outline its underlying assumptions based on the experimental literature. We  then show how the epigenetic competition model can explain paradoxical knockout results in single and multiple knockout experiments in \cite{yun22}.  We then propose a combined model encompassing epigenetic  competition and gene regulation to account for different patterns of gene activity in the observed EMTs. Then, we use our model to offer several predictions, some of which have already been verified. We conclude with a discussion that complements our literature review in the introduction and points to new directions. The mathematical details are included in the supplement.
		
		\section*{Results}
		\subsection*{A new modeling framework accounts for global effects}
		\begin{figure}
			\centering
			\includegraphics[width=1\linewidth]{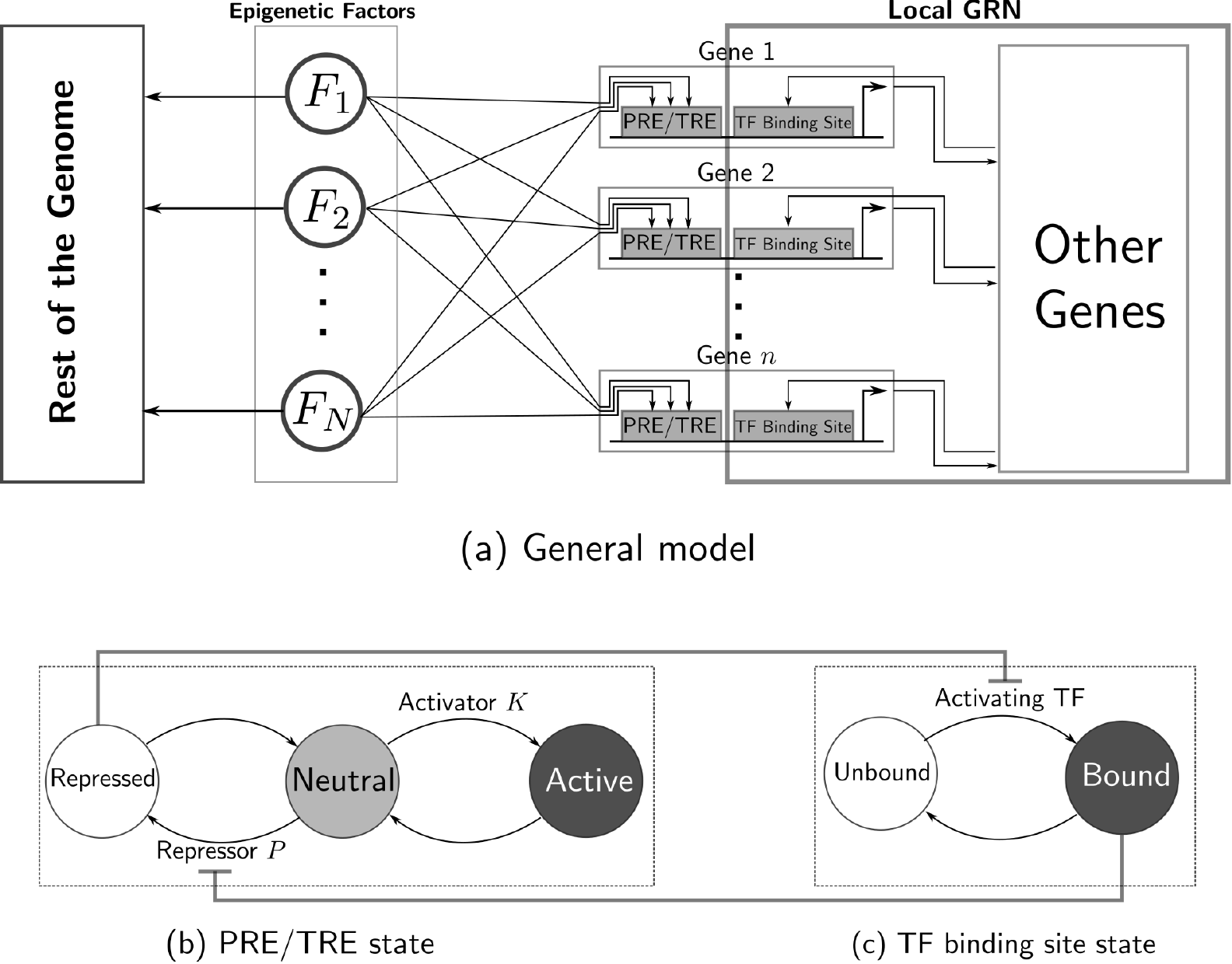}
			\caption{{\bf \sffamily The proposed model.} It consists of three compartments: the local GRN, the EFs, and the rest of the genome. The behavior of the local GRN can be ``tuned'' by global epigenetic factors that compete for similar targets and act genome-wide. On the other hand, the local context of a given GRN affects the binding of global EFs and hence affects their localization. Panels \textbf{(b-c)} Modeling a self-activating gene. (b) The PRE/TRE, (c) The TF binding state. The interaction between EF binding  and TF binding is also illustrated.}
			\label{fig:generalmodel}
		\end{figure}
		
		Consider a GRN consisting of a number of genes interacting with each other via TF binding  and/or micro-RNA-mediated post-transcriptional regulation. The interactions can  either be inhibitory or activating. A subset of genes in the GRN are also subject to the influence of \emph{global} EFs such as histone modifiers (e.g, PcG, TrX). Figure \ref{fig:generalmodel} depicts the schematics of the network. We describe  elements of the model next. The mathematical details are included in the supplement.
		
		\paragraph{Epigenetic Factors}  The EFs can generally be histone modifiers, chromatin remodelers and/or DNA methyltransferases.  We focus here on PcG and TrX EFs, and assume the presence of at least one repressing PcG protein and one activating TrX protein. Based on our review of the literature on the antagonism and redistribution of EFs and epigenetic marks, we make the following four assumptions:
		\begin{enumerate}
			\item \emph{Competition:} The EFs \emph{compete} for binding to similar genomic targets which could be involved in the regulation of the same gene. For example, if a PcG protein binds to a PRE/TRE, or if it deposits a repressive mark, then another TrX protein cannot bind to the same gene, and vice versa. This assumption is justified by our previous review of the literature documenting the antagonism between the two groups of proteins across the genome. 
			\item \emph{Global targeting:} The EFs have targets genome-wide and are not limited to the local GRN. This is justified by the fact that common EFs are known to have hundreds or even thousands of genomic targets.   
			\item \emph{Scarcity:} The levels of the EFs are limited. More formally, we assume that the total levels are constant in the time scale of interest. This is justified by our review of the knockout experiments that observe the dilution of the EFs at their original targets, and their redistribution to new targets.
			\item \emph{Localization:} If an EF is bound to a target on a specific gene, then it cannot simultaneously bind to targets on other genes. This is justified by the observation that EFs (\emph{e.g.}, PRC2) are physically localized to their targets. In addition, the aforementioned dilution and redistribution effects imply that a specific EF complex cannot affect two genes simultaneously. 
		\end{enumerate}
		 {Thus, when an EF is knocked out, there are sufficient binding sites to sequester  the available EFs of another type and make them localize to other genomic  loci.}
		\paragraph{Genes} We use a coarse-grained model of the genes. The individual nucleosomes (that serve as substrates for the various histone-modifying enzymes) are not explicitly modeled. Instead, each gene is modeled as a collection of states that account for the possible histone marks, PRE/TRE occupancies, and TF binding sites. Hence, a given gene can be quantified as distribution of the aforementioned states. The rest of the genome is modeled as a single ``mega-gene'' with a very large copy number compared to the local GRN genes.
		
		\paragraph{Modeling the PREs/TREs} The PRE/TRE component allows the EFs to affect the target gene. For a given gene, the PRE/TRE can be in one the following states: (shown in Figure \ref{fig:generalmodel}-b)
		\begin{enumerate}
			\item \emph{Neutral:} There is no PcG or TrX bound to it, and there are no histone marks.
			\item \emph{Active:} either (a) An activating EF is bound to it (\emph{e.g.}, a TrX protein), or (b) it has an activating histone mark (\emph{e.g.}, H3K27ac or H3K4me3).
			\item \emph{Repressed:} either (a) A repressive EF is bound to it (\emph{e.g.}, a PcG protein), or (b) it has a silencing histone mark (\emph{e.g.}, H3K27me3).
		\end{enumerate}
		Note that the last two states of the PRE/TRE (\emph{Active} and \emph{Repressed}) can have the EF either bound or unbound. The localization effect described earlier arises only for the bound states 2a and 3a. We do not explicitly include a bivalent state as it can be effectively modeled with a gene whose repressed and active states are both present with non-negligible proportions.   
		
		\paragraph{Interaction between TFs and PREs/TREs}
		Our model also allows for interaction between the PRE/TRE state and TF-binding state as shown in Figure \ref{fig:generalmodel}-b,c. This is motivated by the observation that PcG proteins cannot act on genomic loci under active transcription genes, as we have reviewed earlier. The interaction between PRE/TRE and TF-binding states can be modeled by disallowing a PcG protein from silencing a gene while an activating TF is bound to it. In addition, a TF cannot bind to a gene which has been silenced by PcG. Mathematical details concerning the implementation of this effect are provided in SI-\S 2. Note that such an interaction can create regulatory feedback from the local GRN compartment to the EF compartment--- active transcription at the target gene repels repressing EFs.  {Hence, the  {aggregate} effect of  the transcriptional activity of {many} target genes can alter the global level of an EF.}
		
		\subsection*{The genome-wide competition context reverses expected knockout results}
		
		In order to illustrate the model behavior, we consider first a toy example of a single gene that is only regulated by activating and repressing EFs. If the repressive EFs are dominant, then the gene is strongly repressed, while it is strongly active when the activating EFs are dominant. When none of the EFs are present at the gene, we assume that the gene is weakly active. In this scenario, knocking out an activating EF of a gene is expected to reduce expression, while knocking out a repressive EF is expected to increase expression. However, under a competition scenario, opposite effects might occur. We review several cases below.
		
		\begin{figure}
			\centering
			\includegraphics[width=0.95\linewidth]{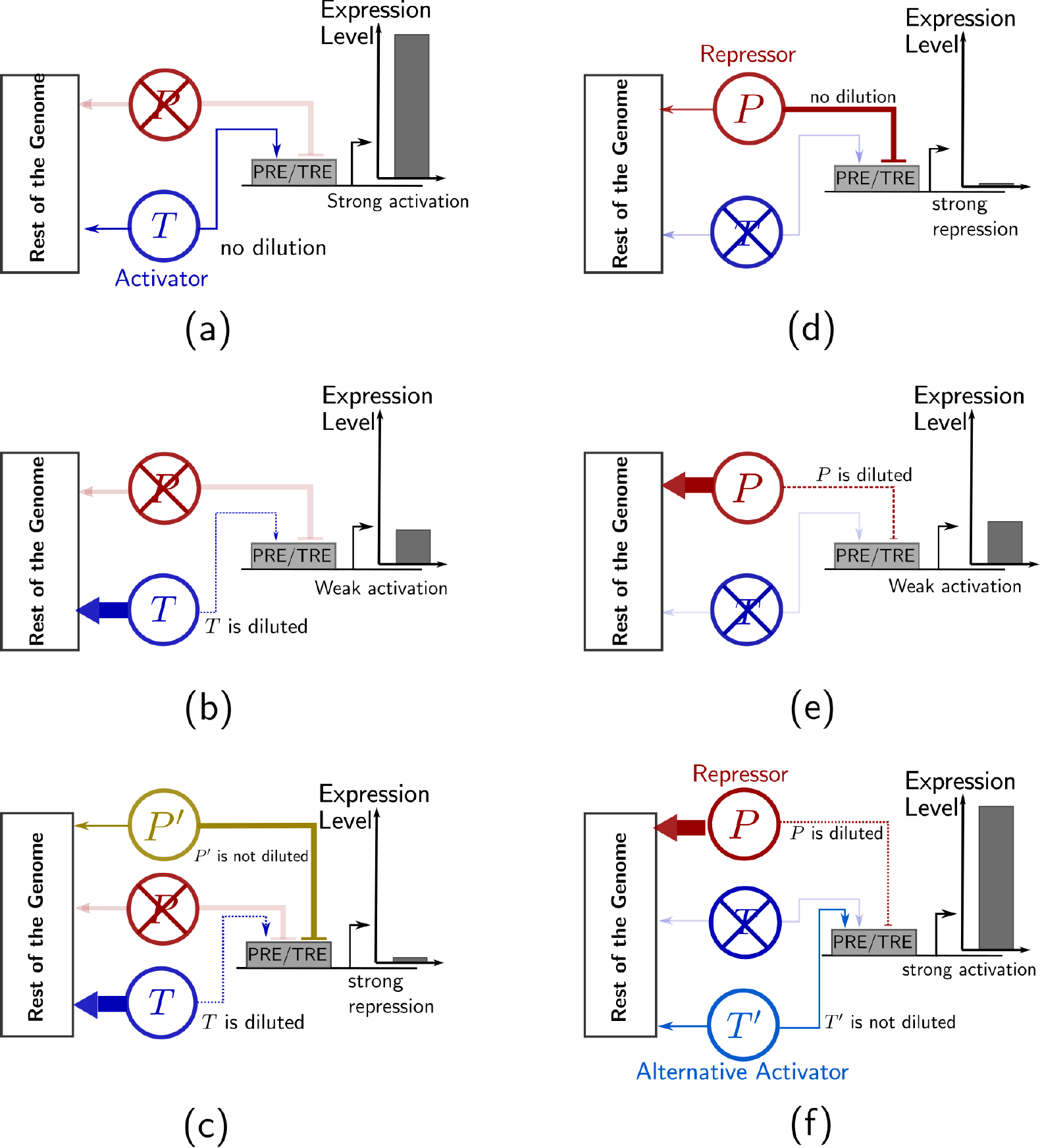}
			\caption{{\bf \sffamily All outcomes are {logically} possible after a knockout experiment due to the competition context.} Panels \textbf{(a)-(c)} {schematically} depict   the impact of knocking-out the repressor $P$, while panels \textbf{(d)-(f)} depict the impact of knocking-out the activator $T$. (a) With minimal competition effects, the EF $T$ can strongly activate its target after its competitor $P$ has been knocked out. (b) The activator $T$ gets diluted after the competitor has been knocked out. Hence, the PRE/TRE becomes mostly unmodified, rendering the gene weakly active.(c) The activator gets diluted, giving an alternative repressor the opportunity to repress the target gene. (d) The EF $P$ represses its target after the competitor $T$ has been knocked out. (e) Knockout of $T$ and dilution of $P$ renders the gene weakly active. (f) The alternative activator $T'$ activates the target gene in the absence of  competition. A very thick arrow denotes sequestration of the corresponding EF by other genes, a dotted arrow depicts dilution of the corresponding EF, and a lightly-shaded arrow denotes an absent regulatory link due to knockout.}
			\label{fig:cases}
		\end{figure}
		
		\paragraph{Two EFs} We first consider one activating EF and one repressive EF as illustrated in Figure \ref{fig:cases}-a,b,d,e. When the competition effects are minimal, knockout of a repressor will induce strong activation of the gene as shown in Figure \ref{fig:cases}-a. Similarly, the knockout of an activator will result in strong repression as shown in Figure \ref{fig:cases}-d. However, when the two EFs compete for targets across the genome, then knocking out one of them can create many new targets for the competing EF. Depending on the EF's binding affinity to the gene under consideration as compared to newly available targets, the competing EF can be diluted genome-wide and redistributed to new targets. This is illustrated in Figure \ref{fig:cases}-b  where knockout of the repressor does not produce strong activation due to the dilution of the activator. Similarly, knockout of the activator does not produce strong repression due to the dilution of the repressor as shown in Figure \ref{fig:cases}-e).
		
		The case with only two EFs is not sufficient to recapitulate all possible outcomes. For instance, it cannot capture a gene that is strongly repressed after the knockout of its only repressor. Such behavior can be recapitulated by a model with three EFs as we show next.
		
		\paragraph{Three EFs} We consider cases with two activators and one repressor, and two repressors and one activator. As reviewed before, there are multiple EFs that have overlapping functions. For instance, H3K4 can be trimethylated by multiple factors. Therefore, the function of a knocked-out EF can be ``rescued'' by an alternative activator. For example, Hanna \emph{et al}.~\cite{hanna22} showed that H3K4me3 levels are elevated at many genomic locations after the knockout of SETD1B (which is an H3K4me3 methyltransferase) due to compensation by MLL2 (KMT2B), an alternative  methyltransferase.  A similar pattern exists for repressors. For instance, it has been observed that the loss of DNA methylation and H3K9 trimethylation is rescued by silencing via H3K27 trimethylation \cite{walter16}.
		
		In order to illustrate our modeling of the aforementioned behavior,  {Figure \ref{fig:cases}-f) shows two activators $T,T'$ and one repressor $P$. 
			Knockout of $T$ and the dilution of the $P$ are not sufficient to explain the strong activation of the gene.  Instead, the presence of an alternate activator $T'$ is required for strong activation. To keep $T'$ undiluted despite the knockout of $T$, the model requires asymmetry between $T'$ and $P$ it terms of their affinity to target sites across the genome.
			Finally, for completeness, we depict in Figure \ref{fig:cases}-f) the case of two repressors and one activator wherein the alternate repressor $P'$ rescues the respression of the gene despite the knockout of the repressing EF $P$. This case corresponds to the experiment reported  in \cite{walter16}. }
		
		\subsection*{The model explains paradoxical knockout results}
		Our modeling framework can explain multiple counter-intuitive results from EF knockout experiments. To illustrate this, we consider the results of knocking out EED (a PRC2 component) and KMT2D (a component of KMT2D-COMPASS complex) in the HMLER cell line \cite{yun22}.  {A} total of 413 genes were identified as targets of PRC2 that   {had significant expression in the control or PRC2-KO cells}. When examining the changes in the expression levels of these genes upon PRC2 and KMT2D knockout, we found multiple genes with paradoxical changes in expression levels. Below, we will use our modeling framework to interpret the observed behaviors.
		
		\begin{table}
			\centering
			\includegraphics[width=\linewidth]{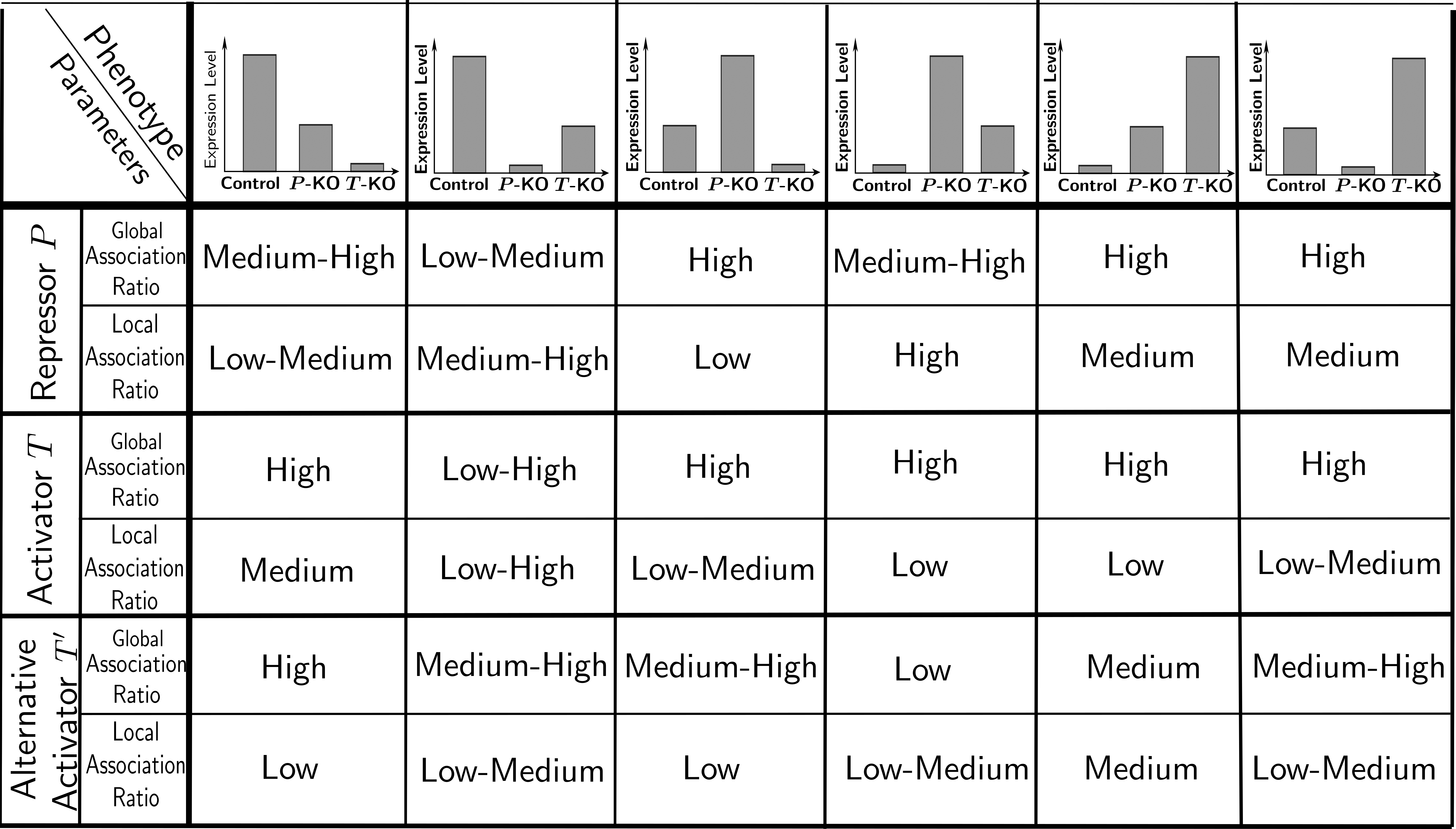}
			\caption{\footnotesize {\bf \sffamily The balance between local and global affinities determines the knockout phenotype.}{\mdseries The  behavior of each EF across the genome is characterized by a mix of \emph{local} and \emph{global} parameters. In this table, we consider a single constitutively-expressed gene. We consider two parameters for each EF: the local association ratio to the gene of interest and the global association ratio to the rest of the genome. We keep the remaining parameters constant. Each entry in the table indicates the average of the corresponding parameter conditioned on the phenotype under consideration. The three experimental scenarios are: Control (both PRC2 and KMT2D are present), $P$-KO (PRC2 is knocked out), and $T$-KO (KMT2D is knocked out). A more detailed version of this table is provided in the Methods section (Table \ref{detailedtable}). }}\label{table}
		\end{table}
		
		{
			\paragraph{The balance between the global and local affinity parameters determines the model's behavior} 
			In order to model the experiments, we study a single constitutively-expressed gene that is subject to the effect of three EFs: one repressing ($P$, e.g, PRC2) and two activating ($T,T'$, e.g., KMT2D and an alternative activator). 
			Consider the three experimental setups: control, $P$ knockout and $T$ knockout. Under such an experiment, we are interested in six possible behavioral phenotypes. Each phenotype is characterized by an unambiguous ordering of the expression levels between the three cases.
			Table \ref{table} shows that all phenotypes are possible if  the global context is considered. In particular, it is shown that the  interplay between global and local affinities of the EFs determines the observed phenotype.

			We illustrate our framework by studying specific examples from the results reported in \cite{yun22}. The majority of the PRC2 target genes (67.71\%) exhibited their highest expression when PRC2 is knocked out, which is expected if the local context is dominant. One such example is CDH2 (N-cadherin), shown in Figure \ref{fig:casesiii}-a. Nevertheless, we still see, unexpectedly, that CDH2 becomes \textit{partially} activated (compared to the control case) when KMT2D is knocked out. This can be interpreted either as PRC2 not being fully diluted upon KMT2D knockout, or due to the weak affinity of the alternate activators (e.g, KMT2C) to CDH2, or a combination of both effects. Next, we study more striking examples. %

			\paragraph{Activator knockout results in strong activation}
			
			In 61 of the PRC2 targets, a paradoxical behavior is reported. The highest expression level was observed when the activating EF KMT2D was knocked out. This includes multiple EMT-related genes such as TWIST1, ZEB1, ZEB2, and PRRX1. Figure \ref{fig:casesiii}-b shows the case of TWIST1 as an example. In the control case, TWIST1 is strongly repressed. When PRC2 (the repressing EF) is knocked out, TWIST1's expression is increased but  {is not strongly activated (when compared to the third case)}, which is counter-intuitive.  {The small magnitude of the increase } in TWIST1 expression upon PRC2 knockout can be explained by   dilution of the activating EFs upon PRC2 knockout.  {Hence, our model's interpretation is that TWIST1 is operating at its nominal level without the presence of EF regulators}. 
			
			 {The third case} is even more surprising where  the knockout of KMT2D (an activating EF) causes TWIST1 to be strongly activated to an expression level that is \textit{multiple times greater} than its expression level when PRC2 is knocked out. As explained in the previous subsection, PRC2 dilution cannot, by itself, explain this paradoxical disparity. This is since PRC2's dilution cannot be worse than a full PRC2 knockout. Mathematically, this implies the existence of an alternative activator $T'$ (\emph{e.g.}, KMT2C) that rescues the expression of TWIST1 when PRC2 is diluted upon KMT2D knockout. This raises the following question: why is the activating effect of $T'$ only observed when KMT2D is knocked out? One possibility within our modeling framework is that $T'$ binds weakly to targets across the genome compared to $P$. Therefore, when $T$ is knocked out, $P$ \emph{out-competes} $T'$ across the genome and it gets diluted. This leaves $T'$ free to activate TWIST1. %
			
			\paragraph{Repressor knockout results in repression}
			
			Another paradoxical behavior can be noticed when examining PRC2 targets that have the highest expression level in the control case. Such genes number 46 out of the 413 PRC2 targets. This set includes CNTN1 (Contactin 1) (Figure \ref{fig:casesiii}-c) which is highly expressed in the control case despite being a target of PRC2. Using our competition paradigm, this can mean that the activating EFs are dominant. Surprisingly, when the repressor PRC2 is knocked out, the expression of CNTN1 is significantly decreased. We interpret this as the outcome of dilution of the activators of CNTN1, caused by the knockout of PRC2. The result is less surprising when KMT2D is knocked out. As seen in Figure \ref{fig:casesiii}-b, loss of PRC2's competitor at CNTN1 allows for stronger repression. {Another possible interpretation of the behavior under our modeling framework is that PRC2 is dominant locally, but its affinity to sites across the genome is much higher compared to the activator. Consequently, PRC2 cannot maintain repression of CNTN1 in the control case. %
		}
		\begin{figure}
			\centering
			\includegraphics[width=1\linewidth]{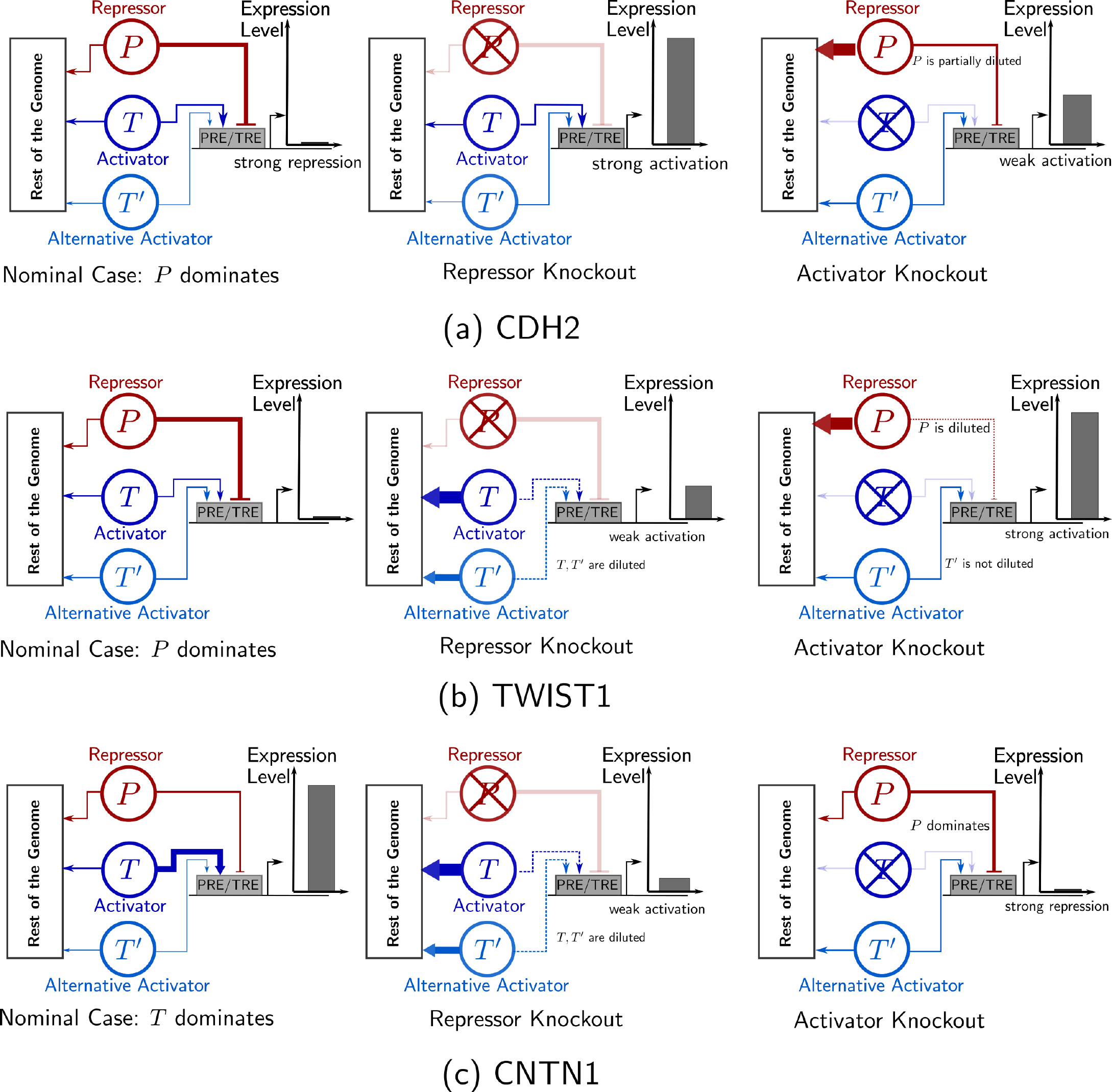}
			\caption{{\bf \sffamily Our model explains the differing behavior of PRC2 targets under two knockout experiments as presented in \cite{yun22}.} Under PRC2 and KMT2D knockouts, the panels show our model's behavior for specific parameter sets that can explain the response of (a) CDH2 (corresponds to the third column in Table \ref{table}), (b) TWIST1 (corresponds to the fifth column in Table \ref{table}), and (c) CNTN1 (corresponds to the first column in Table \ref{table}). A very thick arrow denotes sequestration of the corresponding EF by other genes, a dotted arrow depicts dilution of   the corresponding   EF, and a lightly-shaded arrow denotes an absent regulatory link due to knockout.}
			\label{fig:casesiii}
		\end{figure}

		Overall, the power of our model stems from its versatility and its ability to account for local and global effects simultaneously. In the following sections, we will use our modeling framework to analyze the effect of PRC2 and KMT2D knockout on EMT in HMLER cells \cite{yun22}.
	}

	\subsection*{Global epigenetic factors  modulate the behavior of the local GRN}
	
	\begin{figure}
		\centering
		\includegraphics[width=1\linewidth]{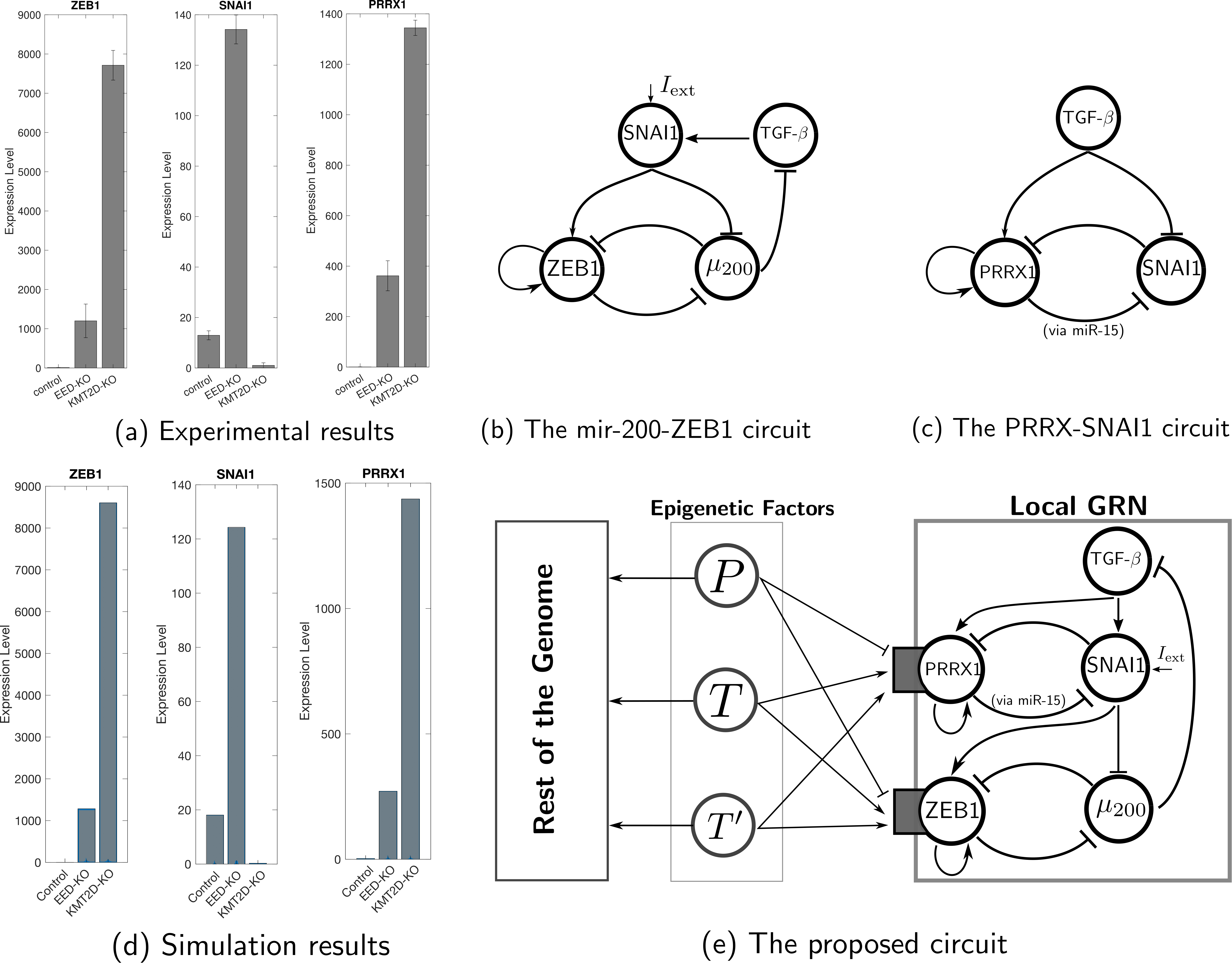}
		\caption{{\bf \sffamily The proposed modeling framework explains the EMT response under EF knockouts.} (a) Experimental RNA-seq results from \cite{yun22}. (b) The miR-200-ZEB1 circuit \cite{lu13}. (c) The PRRX1-SNAIL circuit \cite{fazilaty19}. (d) Model simulations recapitulate the experimental results in panel a. (e) Our proposed model combining ternary epigenetic competition with the mir-200-ZEB1 circuit \cite{lu13,bocci19}, and the PRRX1-SNAIL circuit \cite{fagan19}. $P$ denotes PRC2, $T$ denotes KMT2D, and $T'$ denotes an alternative epigenetic activator such as KMT2C. \ref{fig:review}-a.} 
		\label{fig:review}
	\end{figure}
	
	Zhang \emph{et al}.\ \cite{yun22} characterized the changes in the expression levels of multiple EMT markers in HMLER cells in response to the knockout of two histone methyltransferases, PRC2 and KMT2D. As shown in Figure \ref{fig:review} a, they reported an increase in the mesenchymal regulators ZEB1 and PRRX1 expression levels upon the knockout of both PRC2 and KMT2D, with a much higher fold change upon KMT2D knockout. Thus, both PRC2 and KMT2D knockout resulted in phenotypic change away from the epigenetic state, although to different extents. In contrast to changes in ZEB1 expression, SNAI1 expression as compared to the control case decreased upon KMT2D knockout and was the maximum under PRC2 knockout. This behavior is in disagreement with the traditional transcriptional picture wherein SNAI1 is believed to activate ZEB1 expression, resulting in a positive correlation between their expression levels \cite{lu13}.
	
	The change in ZEB1 and PRRX1 levels in response to PRC2 and KMT2D knockouts can be easily explained using the framework described in the previous section since both of these transcription factors are targets of PRC2 as reported by \cite{yun22}. SNAI1, on the other hand, was not identified as a PRC2 target. To explain the unexpected changes in SNAI1 expression under epigenetic perturbations, we must integrate the transcriptional circuit involving ZEB1, PRRX1, and SNAI1 with our model of EF competition, as described below.

	\paragraph{Integration of EFs competition and local transcription regulation} We consider a system of coupled toggle switches, one involving ZEB1 and miR-200 \cite{lu13} and another involving SNAI1 and PRRX1 \cite{fazilaty19}. As shown in Figure \ref{fig:review} e, the two switches are coupled via the activation of ZEB1 and repression of miR-200 by SNAI1.  {When PRC2 is knocked out, its targets (ZEB1 and PRRX1) are no longer strongly repressed and they get modestly upregulated.  However, they do not get strongly activated due to the dilution of their activators (caused by PRC2 knockout)}. Consequently, the PRRX1-SNAI1 toggle switch will exhibit a high SNAI1 state. However, SNAI1 will be unable to activate ZEB1 in this scenario due to the dilution of ZEB1's epigenetic activator upon PRC2 knockout. When KMT2D is knocked out, PRC2 is diluted at ZEB1 and PRRX1 and hence the alternative activator $T'$ (which can be KMT2C) can fully activate both PRRX1 and ZEB1. The activation of PRRX1 results in the PRRX1-SNAI1 circuit switching to a high PRRX1, low SNAI1 state, in agreement with the experimentally reported behavior. 

	\subsection*{The local transcriptional context determines the effect of the epigenetic factor activity}
	
	In the previous subsection, we have emphasized the effect of EFs on the local GRN. On the other hand, the transcriptional response in a GRN can also be influenced by epigenetic perturbations, mediated by the antagonistic interactions between PcG proteins and active transcription as shown in Figure \ref{fig:generalmodel}-b (see the supplement for detailed models). To demonstrate this effect, we consider the case of a single self-activating gene, here ZEB1 as described below.
	
\begin{figure*}
		\centering
		\includegraphics[width=1\linewidth]{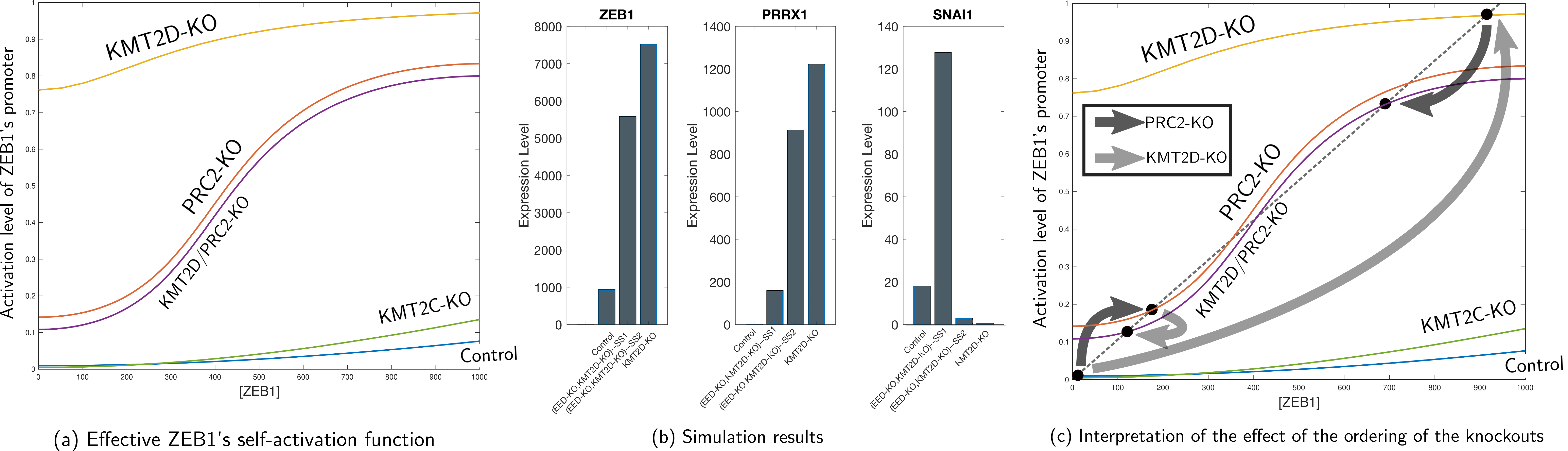}
		\caption{{\bf \sffamily The interplay between the epigenetic and transcriptional contexts}. (a) The self-activation function of ZEB1 for various knockout scenarios, (b) Simulation results of the proposed network in Figure \ref{fig:review}-e showing that the result of PRC2-KO is very different from  KMT2D-KO followed by PRC2-KO. (c) Illustration of the serial knockout experiments using the concept of multi-stability. }
		\label{fig:serialkointerpretation}
	\end{figure*}

	\paragraph{Effect of single EF knockouts} ZEB1 is known to   activate its own promoter \cite{lu13}. Therefore, we study the interaction between the self-activating feedback loop and the EF competition circuit. To that end,  we consider change in the ZEB1 promoter activation level as a function of ZEB1 concentration under the knockout of individual EFs.  The results are pictorially illustrated  in Figure \ref{fig:serialkointerpretation}-a. In the control case (when both PRC2 and KMT2D are present), the activation level of the ZEB1 promoter increases very slowly with the ZEB1 protein concentration due to the inhibitory effect of PRC2. However, if PRC2 is knocked out, we see a sharp, Hill function-like, increase in ZEB1 promoter activation as function of ZEB1 protein concentration. The activation level is low for low ZEB1 levels due to the dilution of KMT2D and other activators away from the ZEB1 promoter upon PRC2 knockout. However, when ZEB1 is abundantly available at its own promoter, the activation level increases due to the self-activatory loop. In the third case, if KMT2D is knocked out instead, the ZEB1 promoter remains highly active even at low concentrations of ZEB1, and there is no substantial change in the activation level of the promoter with ZEB1 concentration. This is because KMT2D knockout is accompanied both by the dilution of repressive PRC2 away from the ZEB1 promoter and strong activity of the alternative activator (e.g, KMT2C)   at ZEB1. Finally, the alternative activator knockout, makes little difference on ZEB1 promoter activation as compared to the control case. Thus, overall, Figure \ref{fig:serialkointerpretation} a shows that epigenetic perturbations do not simply up-regulate or down-regulate their target genes: such interventions can also change the response function of GRN as shown here for the case of a GRN involving a single self-activating gene.

	\paragraph{Effect of double EF knockouts} To further illustrate complexity of the transcription-epigenetic interplay, we next consider the effect of knocking out two EFs in different orders. Figure \ref{fig:serialkointerpretation}-c shows that starting from a GRN state with low ZEB1 expression (phenotypically corresponding to an epithelial state), the GRN will switch to a state with only modestly higher gene expression level (corresponding to a quasi-mesenchymal phenotype). If this is followed by KMT2D knockout, the ZEB1 expression will decrease only slightly. PRC2 knockout, followed by KMT2D knockout, will thus result in epithelial cells switching to a quasi-mesenchymal state. In contrast, if KMT2D is knocked out in epithelial cells, Figure \ref{fig:serialkointerpretation} shows that the cells will switch to a mesenchymal state, one with very high ZEB1 expression. Thereafter, PRC2 knockout will lower the ZEB1 expression only slightly. Thus, double PRC2-KMT2D knockout in epithelial cells will have distinct phenotypic consequences: while PRC2 knockout followed by KMT2D knockout will cause these cells to switch to a quasi-mesenchymal state, KMT2D knockout followed by PRC2 knockout will result in the cells switching to a highly mesenchymal state.

	\subsection*{The modeling framework provides verifiable predictions}
	
	In the previous sections, we have developed a model that integrates the local GRN and global EFs. Using our framework, we can provide several predictions regarding the system studied in \cite{yun22}, as discussed below.

	\paragraph{H3K27me3 redistributes and PRC2 is absent at the promoter of ZEB1 when KMT2D is knocked out} According to our model, an essential mechanism for the activation of ZEB1 and PRRX1 is PRC2's dilution when KMT2D is knocked out. The model predicts a redistribution of PRC2 across the genome resulting in the redistribution of H3K27me3 marks. In addition, our model predicts PRC2's absence at the promoter of ZEB1 and PRRX upon KMT2D knockout. Indeed, it has been observed \cite{yun22} that PRC2 is absent at the promoter of ZEB1 when KMT2D is knocked out, and that H3K27me3 is redistributed to other genomic loci.
	
	\paragraph{EMT does not occur upon the knockout of the alternative epigenetic activator (\emph{e.g.}, KMT2C)} Our proposed model (Figure \ref{fig:review}-e) requires the existence of an alternative activator $T'$. Biologically, $T'$ can correspond to an alternate lysine methyltransferase KMT2C. Using the same parameters used in the simulations depicted in Figure \ref{fig:review}-d), we performed an \emph{in silico} experiment by knocking out the alternative activator $T'$. The resulting behavior is indistinguishable from the control case. In other words, knocking-out $T'$ fails to activate ZEB1 or PRRX1. This is indeed consistent with the screening performed in \cite{yun22} where knocking out KMT2C, for instance, did not result in EMT in HMLER cells.
	
	\paragraph{Simultaneous knockout of KMT2D and PRC2 will not result in strong activation of EMT genes}

	\emph{In silico} experiments using the same parameters used for generating Figure \ref{fig:review}-d show that the result of the simultaneous PRC2-KMT2D double knockout resembles the case of PRC2 knockout far more than the case of KMT2D knockout, \emph{i.e.}, ZEB1 and other EMT genes are not strongly activated under simultaneous PRC2-KMT2D double knockout.%

	\paragraph{Knocking out PRC2 followed by KMT2D will not convert epithelial cells into a highly mesenchymal state}
	
	The activation curves shown in Figure \ref{fig:serialkointerpretation}-a provide us with a predication regarding the cellular response to the PRC2-KMT2D serial knockout experiment: PRC2 knockout will convert epithelial cells to a quasi-mesenchymal state. Thereafter, if PRC2 knockout is followed by KMT2D knockout, Figure \ref{fig:serialkointerpretation}-c shows our model prediction that the cells will stay in a quasi-mesenchymal state and will not switch to a highly mesenchymal state.
	
	\paragraph{Gradual knockouts of PRC2 and KMT2D have different signatures}
	
\begin{figure}
		\centering
		\includegraphics[width=1\linewidth]{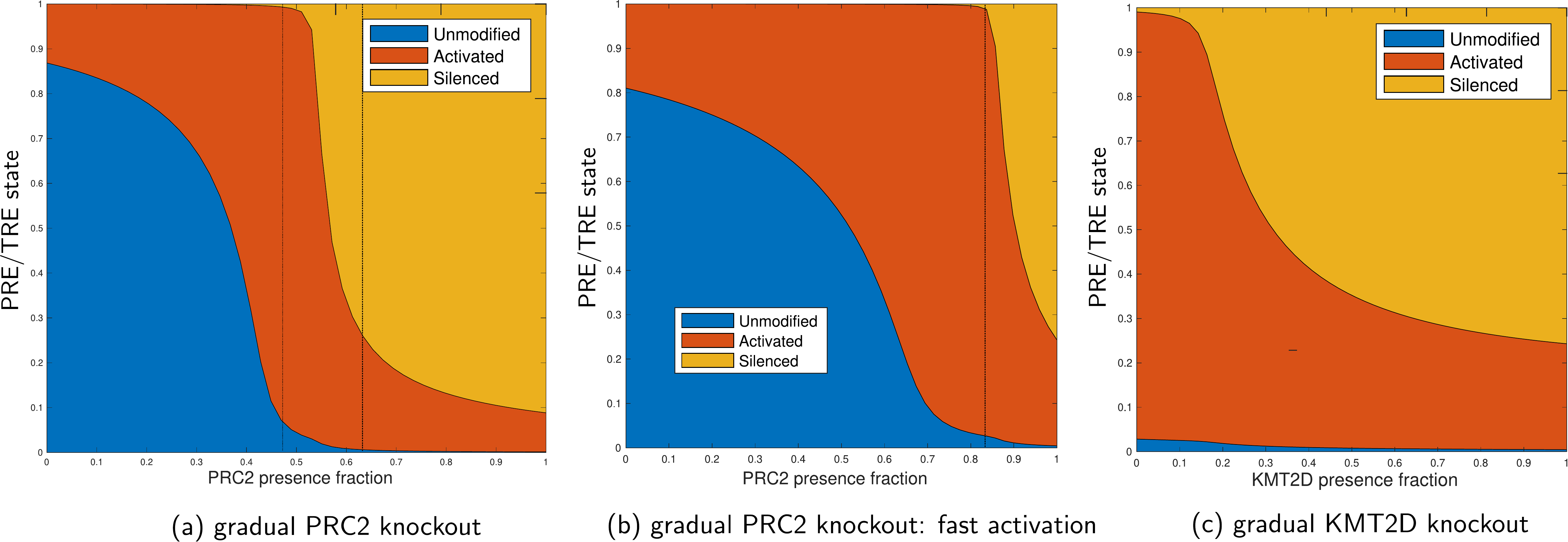} 
		\caption{{\bf \sffamily Predictions of the effect of gradual knockout experiments on ZEB1.} }
		\label{f:predictions_gradual}
	\end{figure}

	Instead of the all-or-none knockout experiments analyzed before, we next consider the case of gradual EF knockouts. Indeed, inhibitors of EZH2 (a sub-component of PRC2) are being investigated as therapy options in cancer \cite{kim16,duan20,eich20}. Figure \ref{f:predictions_gradual}-a shows that ZEB1 is more active when PRC2 is partially knocked out compared to when it is fully knocked out. More precisely, it can be seen that ZEB1 becomes rapidly activated when PRC2's presence fraction goes from 0.6 to 0.5. This rapid activation would indicate the GRN behavior from a regime dominate by PcG to one dominated by TrX. However, when PRC2 is fully depleted, the levels of TrX get gradually diluted away from the ZEB1 promoter, leaving the PRE/TRE of ZEB1 in an unmodified state and weakly active.
	
	Our model would suggest that this effect can be more pronounced: simulations with different model parameters, as shown in Figure \ref{f:predictions_gradual}-b, would indicate the possibility of a situation where a small dip in the level of PRC2 can cause a rapid collapse in its activity at the ZEB1 promoter. This can be contrasted with the mechanism of activation in the case of a gradual KMT2D knockout. In that scenario, the activity of the ZEB1 PRE/TRE builds up \emph{slowly} as PRC2 gets increasingly diluted and as the alternative epigenetic activator gets the full chance to activate ZEB1 as shown in Figure \ref{f:predictions_gradual}-c. From these results, we can make the counter-intuitive prediction that that the partial knockout of an epigenetic repressor, here PRC2, can have a stronger repressive effect on gene activity as compared to a complete PRC2 knockout.
	
	\section*{Discussion}
	
	In this work, we have described a new modeling framework that combines local transcriptional regulation with global epigenetic control, and showed that complex interplay between transcriptional and epigenetic control can lead to rich gene expression dynamics. We have used our modeling framework to understand the effects of various epigenetic perturbations on epithelial-mesenchymal transition, a crucial cellular process involved in both health and disease. We note that interplay between epigenetic competition, the EMT transcriptional network, and the baseline transcriptional context can result in counter-intuitive experimental observations, and generate unique paths for cells to transition between epithelial and mesenchymal states.
	 {Furthermore, our results indicate that the rest of the genome exerts an indirect effect on the behavior of the local GRN by competing for the same EFs.  So when one of the EFs is knocked out, the new empty sites across the rest of the genome try to sequester the other EFs present at the local GRN As a result, the local landscape is reshaped  leading to new steady states. A key property of the model  is the asymmetry between the behaviors of different EFs as shown in Table \ref{table}. In other words, the number of global sites that are made available to a specific EF depends on the identity of the eliminated and the competing EFs, and the local and global affinities of the considered EFs. }
	
	\subsubsection*{\upshape Further evidence of redistribution and dilution of PcG proteins} Our modeling framework relies upon the competition between PcG and TrX proteins to modify histones at the same or nearby genomic sites, and the redistribution of PcG that accompanies TrX knockout. Such an effect is not just restricted to TrX and PcG proteins. Another line of research has investigated the relationship between  DNA methylation, H3K9me3, and H3K27me3. Despite the fact that all three are silencing marks, they do not usually mark the same genomic locations. DNA methylation and H3K9me3 mark constitutive heterochromatin, while H3K27me3 marks facultative heterochromatin \cite{deleris12, basenko15}. Induced DNA hypo-methylation can cause H3K27me3 to disappear from its normal locations and accumulate at new genomic sites in \textit{Arabidopsis thaliana} \cite{deleris12},  mouse somatic cells \cite{reddington13}, and mouse ESCs \cite{Douillet20}. Hence, inhibition of DNA demethylation can open up new locations for PRC2 recruitment, sequestering it from PcG-repressed genes \cite{Douillet20}. Similarly, elimination of H3K9me3 has been shown to cause H3K27me3 to disappear from its normal genomic locations and redistribute to new genomic loci in the fungus \textit{Neurospora} \cite{basenko15, jamieson16}. It is worth noting that PRC2 sequestration and redistribution is not limited to competition scenarios. For instance, it has been observed that PRC2 redistributes across the genome upon modifying ATRX, a chromatin remodeler that assists PRC2's binding \cite{ren20}.
	
	\subsubsection*{\upshape Disorder in PcG/TrX proteins causes disease and is a target for therapeutics}
	
	Given the global activity of PcG and TrX proteins, it is expected that their knockout will have far-reaching and detrimental effects on cells. Indeed, the inactivation or aberrant activation of such proteins has been shown to play a key role in the emergence of cancer \cite{morgan15}. PRC2  has been studied extensively in that context \cite{laugesen16,dockerill21}, and its catalytic component EZH2 has been tested as a therapeutic target in multiple clinical trials \cite{kim16,duan20,eich20}. Similarly, disorders in COMPASS proteins are very common in cancers \cite{sze16,fagan19}, and have been proposed as key regulators and potential therapeutic targets \cite{dawkins16,gala18,xiong19,yu20,dhar21}. Our results imply that such thereputic interventions must proceed with the utmost caution by accounting for the global context.
	
	\subsubsection*{\upshape Similar EFs can play different roles} 
	Our results show that EFs with the same enymatic activity, such as KMT2C/D both of which deposit the same methylation mark on histone tails, can exhibit very different biological and functional behaviors. This has been reported in multiple experimental contexts. For example, Zhang \emph{et al}.\ found that while KMT2D knockout could induce transition to a mesenchymal state in HMLER cells, KMT2C was not identified to be among the key EMT inducers \cite{yun22}. Similarly, in MCF10A cells, TGF-$\beta$ induced EMT is accompanied by the upregulation of the H3K27me3 demethylase KDM6B while the enzymatically similar KDM6A is downregulated during the process \cite{mani17}.  
	
	\subsubsection*{\upshape Competition effects in molecular biology} 
	Competition effects have been studied earlier in the context of synthetic biological circuits where circuits compete for RNA polymerases and ribosomes. It has been shown that such competition can cause nontrivial coupling between isolated components and affect protein expression performance \cite{qian17,MA_LCSS22}. It has also been investigated in the context in the design of Boolean genetic circuits via CRISPRi which uses dCas9 as a shared resource \cite{shuyi18}. More detailed models of mRNA's competition for ribosomes during the transcription process have also been proposed \cite{raveh16,miller21}. Similarly, competition for the same micro-RNAs between transcripts that have the same or similar micro-RNA-binding sequence motifs has been shown to induce coupling between the expression levels seemingly independent proteins \cite{bosia, salmena}.
	
	\subsubsection*{\upshape Experiments involving epigenetic perturbations must be analyzed with care} One crucial takeaway from the analysis presented in this manuscript is the possibility of widespread cross-talk between the genomic targets of different epigenetic factors. Most experimental studies analyzing the effect of epigenetic perturbations follow a set procedure: characterize the transcription profiles before and after knocking out an epigenetic modifier, identify the set of differently-expressed genes (which often number in the hundreds), and carry-out gene set enrichment analysis \cite{gsea} or Gene Ontology enrichment analysis \cite{ontology} using the differently-expressed genes. In light of the framework described here, it is unsurprising that the outcome of epigenetic perturbations is hundreds of differently-expressed genes, a list which is then arbitrarily whittled down depending on the biological interests of the researchers carrying out the analysis. This is usually followed by choosing a pathway or biological process of interest and analyzing how it is affected by the given epigenetic modifier. One would then conclude with identifying that epigenetic modifier as a key regulator of that biological process.  {Our analysis shows that any such conclusion could be highly unreliable outside the context of the specific experimental setup. For example, in the study by Zhang \emph{et al.} \cite{yun22}, SNAI1 expression is upregulated upon PRC2 knockout even though it is not a direct target of PRC2. While a simple differential gene expression analysis might lead one to identify PRC2 as a key regulator of SNAI1 expression, our analysis shows that the effect of PRC2 on SNAI1 expression can only be explained by the complex interplay between epigenetic control and the GRN involving SNAI1, ZEB1, and PRRX1. Thus, PRC2 knockout may have no effect on SNAI1 expression in cells wherein the SNAI1-ZEB1-PRRX1 GRN is inactive. Moreover, we show that the effect of epigenetic perturbations on gene expression will depend on the cell's transcriptional state at the time of the epigenetic perturbation. This would suggest that the same epigenetic perturbation could have very different effects on genotypically identical cells in different phenotypic states. For example, PRC2 knockout has been shown to trigger differentiation in primed mouse embryonic stem cells but not in naive embryonic stem cells \cite{shanprc2}. Thus, careful analysis of the effects of epigenetic perturbations will also require that the effects be analyzed within the transcriptional context of the cells.}
	\section*{Methods}
	{\vspace{-0.25in}{\subsection*{Data sets} The data sets used in this study include the full RNAseq data sheet and the list of PRC2 targets for the experiments in reported \cite{yun22}. The data has been provided by Yun Zhang who is the lead author in the aforementioned study.

			\subsection*{Mathematical modeling}
			The details of the mathematical models and numerical simulations are provided in the supplement, and are very briefly summarized here. 
			\subsubsection*{Constitutively-expressed genes subject to EF competition} In SI-\S1 we describe the reaction network models for a general network of $N$ EFs and $n$ genes. We show that each interaction of an EF with a gene can be characterized by two parameters: the association ratio and the marking ratio. In particular, the results in Table \ref{table} are generated by studying the effect of the parameters on the three experimental scenarios discussed in \S C earlier. A more detailed version of Table \ref{table} is shown in Table \ref{detailedtable}.
			\begin{table*}
				\centering
				\includegraphics[width=\linewidth]{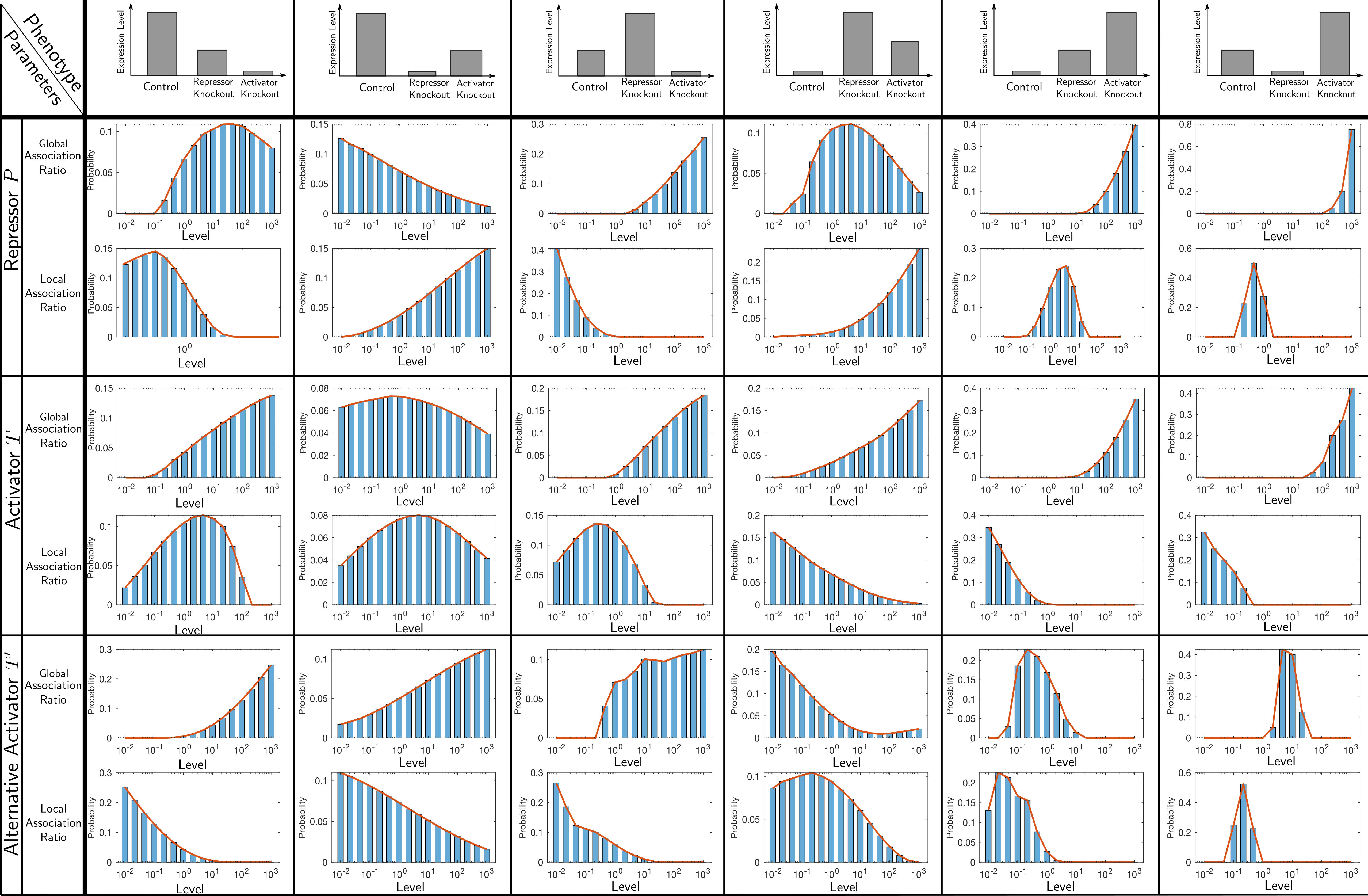}
				\caption{\footnotesize\textbf{All  phenotypes  are possible under the appropriate mix of global and local contexts.}{\mdseries  This is a more detailed version of Table \ref{table}. Each plot depicts the marginal probability distribution of the parameter under consideration conditioned on the phenotype under consideration.  The marking ratios are fixed, while the association ratios are varied.  The six association ratios are varied with 16 levels between $10^{-3}$ and $10^3$. This provides $6^{16}\approx 2.8211\times 10^{12}$ sets of parameters. For each of which, the steady states are calculated numerically for the control and the knockout cases. A set of parameters is said to give one of the six phenotypes    if the highest expression level (amongst the three cases) is at least three times higher than the second highest, and latter is at least three times  higher than the third highest. To generate this table, the total $P,T,T'$ levels are $1000,500,100$, respectively. The local marking ratios are fixed to 1, while the global marking ratios are fixed to 0.01. The total copy number of the local gene is fixed to 1, while the total copy number of the rest of the genome is fixed to 1000. }}\label{detailedtable}
			\end{table*}
			
			\subsubsection*{Self-activating genes subject to EF competition} In SI-\S 2, we provide our mathematical model for the interaction between the transcriptional and epigenetic components of regulation for a group of $n$ self-activating genes and $N$ EFs. We illustrate this numerically by showing the effect of EF perturbations on the activation function of a single self-activating gene.
			
			\subsubsection*{A general local GRN subject to EF competition} In SI-\S 3, we model a generic local GRN with TF and micro-RNA regulations subject to EF competition. We provide our concrete model for the network in Figure \ref{fig:review}-e along with the parameters utilized in the simulations.

			\subsubsection*{Parameter selection}  {The feasible parameters are not unique. The simulation parameters have been chosen to reproduce the qualitative behavior of the experimental results.   For    the epigenetic competition circuit, the parameters have been chosen based on  a screen similar to those shown in Table \ref{detailedtable} and Supplementary Tables 1-3.  The parameters of the local GRN have been chosen by refining an initial parameter set generated by the software package RACIPE \cite{racipe}. The parameters are listed in the supplement. 
			}
			
			\subsection*{Software} Numerical simulations  have been performed via MATLAB R2020a on the discovery HPC cluster at Northeastern University. 
			
	}}

	\paragraph{Acknowledgment}{M.A.R, S.T, and H.L were supported by the National Science Foundation grant PHY-2019745. M.A.R and E.S were supported in part by grant NSF/DMS-2052455. E.S was supported in part by AFOSR FA9550-21-1-0289 and AFOSR 22RT0159.
		Y.Z. was supported by Susan G. Komen Postdoctoral Fellowship no. PDF15301255. We thank Robert A. Weinberg for useful conversations and his support of the work of Y.Z. We also acknowledge fruitful discussions with Sendurai Mani.}

	\newpage 
	\renewcommand{\figurename}{{\bfseries Supplementary Figure}}
	\renewcommand{\tablename}{{\bfseries Supplementary Table}}
\setcounter{figure}{0}
\setcounter{table}{0}
	\begin{center}
		{\Large \sffamily \bfseries Supplement to \\ ``Epigenetic factor competition reshapes the  EMT landscape''}
	\end{center}

	\noindent This supplement contains the details of the mathematical models utilized, additional numerical simulations, and the parameters used.
	\subsection*{Notation}
	We use the formalism of Biological Interaction Networks (BIN) (also known as, Chemical Reaction Networks) as expounded in \cite{feinberg87,erdi89,MA_LEARN}. We review basic notations below.
	
	A reaction network consists of \emph{species} $\mathscr S=\{\mr X_1,..,\mr X_n\}$ and \emph{reactions} $\mathscr R=\{\mathrm R_1,..,\mathrm R_\nu\}$. Examples of species include binding sites, PRE/TREs, mRNAs, proteins, etc. Examples of reactions include binding, unbinding, dimerization, production, decay, etc. More formally, the reaction $\rm R_j$ is written in the following form:
	$ \sum_{i=1}^n \alpha_{ij}   \mr X_i \longrightarrow \sum_{i=1}^n \beta_{ij} \mr X_i, $
	where $\alpha_{ij},\beta_{ij}$ are nonnegative integers. 
	The stoichiometry matrix of the network is defined element-wise as $[\Gamma]_{ij}=\beta_{ij}-\alpha_{ij}$ and it describes the net gain or loss of the $i$th species at the $j$th reaction.
	
	Each species $\mr X_i$ is a quantified by a concentration $X_i \ge 0$, while a reaction $\rm R_j$ is quantified by a reaction rate or velocity function $V_j$. We use the standard Mass-Action kinetics written as follows: 
	$V_j(x)=\prod_{i=1}^n k_j x_i^{\alpha_{ij}}$, where $k_j's$ are the kinetic constants.
	
	In order to describe the time-evolution of the network, the corresponding Ordinary Differential Equation (ODE)   can be written as follows:
	\begin{equation}\label{e.ode} \dot X = \Gamma V(X), X(0)=X_\circ. \end{equation}
	
	BINs usually have conserved quantities. In our work, we assume that the genes and the Epigenetic Factors (EFs) are conserved. Mathematically, this means that there exists a nonnegative vector $d \in \mathbb R^n$ such that $d^T \Gamma=0$. In addition $d^T X(t) =d^T X(0)= \mbox{constant}$ for all $t\ge 0$.
	
	In this paper, we are mostly interested in steady-state analysis. Therefore, we will  solve the following system of equations:
	\begin{equation}\label{e.ss}
		0=\Gamma V(X), d_j^T X = c_{k,total},k=1,..,m,
	\end{equation}
	where $d_j, c_{k,total}, k=1,..,m$ are the associated conservation laws and conserved quantities, respectively.
	
	\section{Modeling a single gene subject to epigenetic factor competition}
	
	\subsection{Basic model}
	As shown in Figure 	1, the network has $N$ EFs $\mr F_1,..,\mr F_N$. The local GRN has $n$ genes that are subject to the effect of the EFs. In addition, the EFs affect the rest of the genome which is modeled as a single ``mega-gene'' and we give it the index $0$. Hence, the EFs have a total of $n+1$ targets. 
	
	\paragraph{Competition network} Let consider the $i$th gene, with $i  \in \{0,1,..,n\}$. Denote the corresponding PRE/TRE component  by $  \mr G\s{i}$. Then, we assume it can have the following states: 
	\begin{enumerate}
		\item \emph{Unbound:} It is denoted by $\mr G_{0}\s{i}$ which means that nothing is bound to $\mr G\s{i}$ and there are no histone marks.
		\item \emph{Bound by the EF $\mr F_j$:} It is denoted by $\mr G_{j}\s{i}$. The $j$th factor is bound to $\mr G_i$. For simplicity, we assume that the EF marks the corresponding histone immediately.
		\item \emph{Unbound and marked:} We denoted it either by $\mr G_{+}\s{i}$ or $\mr G_{-}\s{i}$ depending if the histone mark is activating or repressing, respectively. 
	\end{enumerate}
	Therefore, we can write the following  of reactions for a repressing EF $\mr F_j$:
	\begin{align}\label{compCRN1}
		\mr F_j+\mr G_{0}\s{i}  \xrightleftharpoons[a_{-ij}]{a_{+ij}} \mr G_{j}\s{i} \lra^{\gamma_{+ij} } \mr G_{-}\s{i}+\mr F_j, ~ \mr G_{-}\s{i}  \lra^{\gamma_{-i}} \mr G_{0}\s{i}. 	
	\end{align}
	Similarly, we write the following for an activating EF $\mr F_j$:
	\begin{align} \label{compCRN2}
		\mr F_j+\mr G_{0}\s{i}  \xrightleftharpoons[a_{-ij}]{a_{+ij}} \mr G_{j}\s{i} \lra^{\gamma_{+ij}} \mr G_{+}\s{i}+\mr F_j, ~ \mr G_{+}\s{i}  \lra^{\gamma_{+i}} \mr G_{0}\s{i}. 	
	\end{align}
	In plain words, the reactions describe the binding/unbinding of $\mr F_j$ to the PRE/TRE $\mr G_{0}\s{i}$, and the marking of the corresponding histones. The histone mark can be erased either constitutively, or via the activity of histone modifiers which are not explicitly modeled. 
	The overall network has $N+(n+1)(N+3)$ species and $4N(n+1)$ reactions.
	
	The network above models the four assumptions postulated in the main text. \textit{Global-targeting} is captured by the PRE/TRE of the mega-gene denoted by $\mr G\s{0}$. In our simulations, we set $G_{total}\s0 \gg G_{total}\s1$ to model the fact that the total number of targets across the genomes is large. \textit{Competition} is captured by the fact that two different EFs can not bind to a specific PRE/TRE $\mr G\s{i}$ simultaneously. \emph{Localization} is captured by the fact that the bound EFs $\{\mr G_{j}\s{i}\}$ can not interact with other genes. Finally, \textit{scarcity} is captured by the stoichiometric conservation of the EFs. In other words, there are no new EF molecules that are created or annihilated in the network above. More concretely, for each $j=1,..,N$, we have \begin{equation} \label{Fconv} F_j+\sum_{i=0}^n   G_{j}\s{i} =   F_{j,total}.\end{equation}

	The total copy number of each gene is fixed, hence, the PRE/TREs are also conserved. In particular, we have for each $i=0,..,n$: \begin{equation}\label{e.Gconv}  G_0\s{i} +   G_+\s{i} +   G_-\s{i} +  \sum_{j=1}^N   G_{j}\s{i} =  G_{total}\s i,\end{equation} where $  G_{total}\s i$ is equal to total concentration of the $i$th gene. Therefore, at each point of time, we can view the epigenetic state of the $i$th gene as a distribution of $N+3$ states that span the unbound,  positively and negatively marked, and the bound. %
	
	\paragraph{Steady-state expressions} By writing the ODEs for the competition model above, the steady-state values of the PRE/TREs states can be found. To that end, let us partition the EFs into   the activating and repressing subsets:  $J_+=\{j \in \{1,..,N\}| \mr F_j~\mbox{is activating} \}$, $J_-=\{j \in \{1,..,N\}| \mr F_j~\mbox{is repressing} \}$. The interaction between a particular EF and gene pair can be characterized by two effective parameters:
	\begin{enumerate}
		\item The \textit{EF association ratio} of $\mr F_j$ to $\mr G_j\s{i}$ is defined as $a_{ij}:=a_{+ij}/(a_{-ij}+\gamma_{+ij})$. 
		\item The \textit{histone marking ratio}, and is defined as $\gamma_{ij}:=\gamma_{+ij}/\gamma_{+i}\s{i}$ if $j\in J_+$, and $\gamma_{ij}:=\gamma_{+ij}/\gamma_{-i}$ if $j\in J_-$.
	\end{enumerate}
	Therefore, using \eqref{e.Gconv}, we can write:	
	\begin{align} \label{G} G_{j}\s{i}&=G_{tot}\s{i}F_j \frac{ a_{ij} }{1+\sum_{j=1}^N c_{ij} F_j},&G_{-i} &=G_{tot}\s{i}\sum_{j\in J_-} F_j \frac{ a_{ij}   \gamma_{ij}  }{1+\sum_{j=1}^N c_{ij} F_j} \\ \nonumber G_0\s{i}&= G_{tot}\s{i} \frac{1}{1+\sum_{j=1}^N c_{ij} F_j}   , & G_{+}\s{i}&=G_{tot}\s{i} \sum_{j\in J_+} F_j \frac{ a_{ij}   \gamma_{ij}  }{1+\sum_{j=1}^N c_{ij} F_j} .  \end{align}
	where  $c_{ij}:=a_{ij}(1+\gamma_{ij})$ is the \textit{regulation ratio} which depends on the two aforementioned parameters.%
	
	By examining the expressions above, it can be seen that the different genes are only coupled via the the 
	EFs $F_1,..,F_N$.
	Unlike the neat expressions above, determining the free EFs   requires solving \eqref{Fconv}. By substituting from \eqref{G}, we get 
	\begin{equation} \left (1+ \sum_{i=1}^n \frac{G_{tot}\s{i} a_{ij} }{1+\sum_{j=1}c_{ij} F_j } \right ) F_j = F_{j,total},\end{equation}
	which yields the  polynomial equation:
	\begin{equation}\label{polynomial} \prod_{i=1}^n \left ( 1+\sum_{j=1}c_{ij} F_j\right ) (F_j-F_{j,total}) + \sum_{i=1}^n G_{tot}\s{i} a_{ij} F_j \prod_{\tilde i\ne i}  \left ( 1+\sum_{j=1}c_{\tilde ij} F_j\right )=0. \end{equation}
	
	Hence, in order to find the free levels of the EFs we need to solve a system of coupled $(n+2)$th-order polynomials. Even with a single local gene and two factors, this amounts to two coupled  cubic equations which are infeasible to solve analytically. Luckily, it can be shown that the reaction network \eqref{compCRN1}-\eqref{compCRN2} is always injective \cite{craciun05}, hence we can state the following result which can be proved by showing that the Jacobian is always $P_0$ \cite{banaji07}:
	\begin{theorem} Let the EFs $F_1,..,F_N$, and genes $\mr G\s1, ..., \mr G\s n$ be given. For any fixed total EFs $F_{1,tot},..,F_{N,tot}$ and total genes $G_{tot}\s1,...,G_{tot}\s n$, the reaction network \eqref{compCRN1}-\eqref{compCRN2} can not admit more than a single positive steady.
	\end{theorem}
	In our simulations, we recourse to numerical methods (Newton-Raphson or ODE solvers) to evaluate the unique solution of the system \eqref{polynomial}.
	
	\paragraph{Gene expression network} To model the manner in which the epigenetics affect transcription, we assume that the expression is most active when either an activating EF is bound or if the histone is positively marked. Furthermore, we assume a small residual expression when the gene is epigenetically unmodified, and zero expression when the gene is silenced. This can be written as the following network for the $i$th gene:
	\begin{align} \nonumber
		\mr G_{0}\s{i} &\lra^{k_+} \mr G_{j}\s{i}+\mr X\s{i}, j \in J_+, \\ \mr G_{+}\s{i} &\lra^{k_+} \mr G_{+}\s{i}+\mr X\s{i}, \\  \nonumber
		\mr G_i\s{0} &\lra^{\rho k_+} \mr G_i\s{0}+\mr X\s{i}, \\\nonumber
		\mr X\s{i} & \lra^{k_-}  \emptyset.
	\end{align}
	where $0<\rho < 1$ is the residual expression ratio of the unmodified state.
	Therefore,  the steady state expression level is given as $X_i=\frac{k_+}{k_-} \Psi_i$, where $\Psi_i$ is the epigenetic activation function of the $i$th gene written as:
	\begin{equation}\label{Psi1} \Psi_{i}=\sum_{j \in J_+}   G_{j}\s{i} +   G_{+}\s{i} + \rho   G_i\s{0}=G_{tot}\s i\frac{ \rho +\sum_{j\in J_+}c_{ij} F_j  }{ 1+\sum_{j=1}c_{ij} F_j},\end{equation} 
	and $F_1,..,F_N$ are the solutions of \eqref{polynomial}.

	\subsection{Examples}
	
	In order to find the activation functions, we will give few examples below. 
	
	\paragraph{A single gene and two EFs} As in the main text, we consider a toy example of a single gene $\mr G_1$ subject to the effect of two EFs: one activating (denoted by $T$) and one repressing (denoted by $P$). Recall that the rest of the genome is denoted by $\mr G\s0$. Therefore, the   BIN describing the system is given as follows:
	\begin{align} \label{e.2EF1}
		{\tiny\begin{array}{c}\mbox{Target}\\\mbox{Gene}\end{array}} &\left \{\begin{array}{rl}\mr T+\mr G_{0}\s{1} \hspace{-0.1in} & \rightleftharpoons \mr G_{T}\s{1} \lra \mr G_{+}\s{1}+ \mr  T, ~ \mr G_{+}\s{1}  \lra \mr G_{0}\s{1} \\
			\mr  P+\mr G_{0}\s{1}  \hspace{-0.1in} & \rightleftharpoons \mr G_{P}\s{1} \lra \mr G_{-}\s{1}+ \mr  P, ~ \mr G_{-}\s{1}  \lra \mr G_{0}\s{1} \end{array} \right . \\ \label{e.2EF0}
		{\tiny\begin{array}{c}\mbox{Rest of}\\\mbox{the Genome}\end{array}}&\left\{\begin{array}{rl}\mr  T+\mr G_{0}\s{0} \hspace{-0.1in} & \rightleftharpoons \mr G_{T}\s{0} \lra \mr G_{+}\s{0}+\mr  T, ~ \mr G_{+}\s{0}  \lra \mr G_{0}\s{0} \\
			\mr  P+\mr G_{0}\s{0}  \hspace{-0.1in} & \rightleftharpoons \mr G_{P}\s{0} \lra \mr G_{-}\s{0}+\mr  P, ~ \mr G_{-}\s{0}  \lra \mr G_{0}\s{0} \end{array} \right..
	\end{align}
	Let $a_{1T},a_{1P},a_{0T},a_{0P},\gamma_{1T},\gamma_{1P},\gamma_{0T},\gamma_{0P}$ be the EF association and histone marking ratios, respectively. As before,  let  $c_{1T}= a_{1T}(1+\gamma_{1T}), c_{1P}= a_{1P} (1+\gamma_{1P}),c_{0T}= a_{0T}(1+\gamma_{0T}), c_{0P}= a_{0P} (1+\gamma_{0P}) $.

	By solving the associated equation \eqref{e.ss},  the activation function takes the following form:
	\begin{equation}\label{ActiveE} E=    G_{tot}\s{1}\frac{\rho  + c_{1T} T  }{1+ c_{1T} T +c_{1P} P } .
	\end{equation}

	In order to find $P,T$, we need to solve \eqref{polynomial} which can be written as a pair of cubic equations:
	{\scriptsize	\begin{align*}
			0&=(T - T_T)(1+c_{1T}T+c_{1P}P)(1+c_{0T}T+c_{0P}P)+ G_{1,tot} a_{1T} T (1+c_{0T}T+c_{0P}P) + G_{0,tot} a_{0T} T(1+c_{1T}T+c_{1P}P)\\
			0&=(P - P_T)(1+c_{1T}T+c_{1P}P)(1+c_{0T}T+c_{0P}P)+ G_{1,tot} a_{1P} P (1+c_{0T}T+c_{0P}P) + G_{0,tot} a_{0P} P(1+c_{1T}T+c_{1P}P)
	\end{align*}}
	As in the main text, we assume that the gene is only regulated by the PRE/TRE. Let the X be the protein expressed. Hence, we write $ \emptyset \xrightleftharpoons[k_-]{k_+ E }\mr  X$. Therefore, we get $X= (k_+/k_-)E $.

	\begin{figure}
		\centering
		\includegraphics[width=0.9\linewidth]{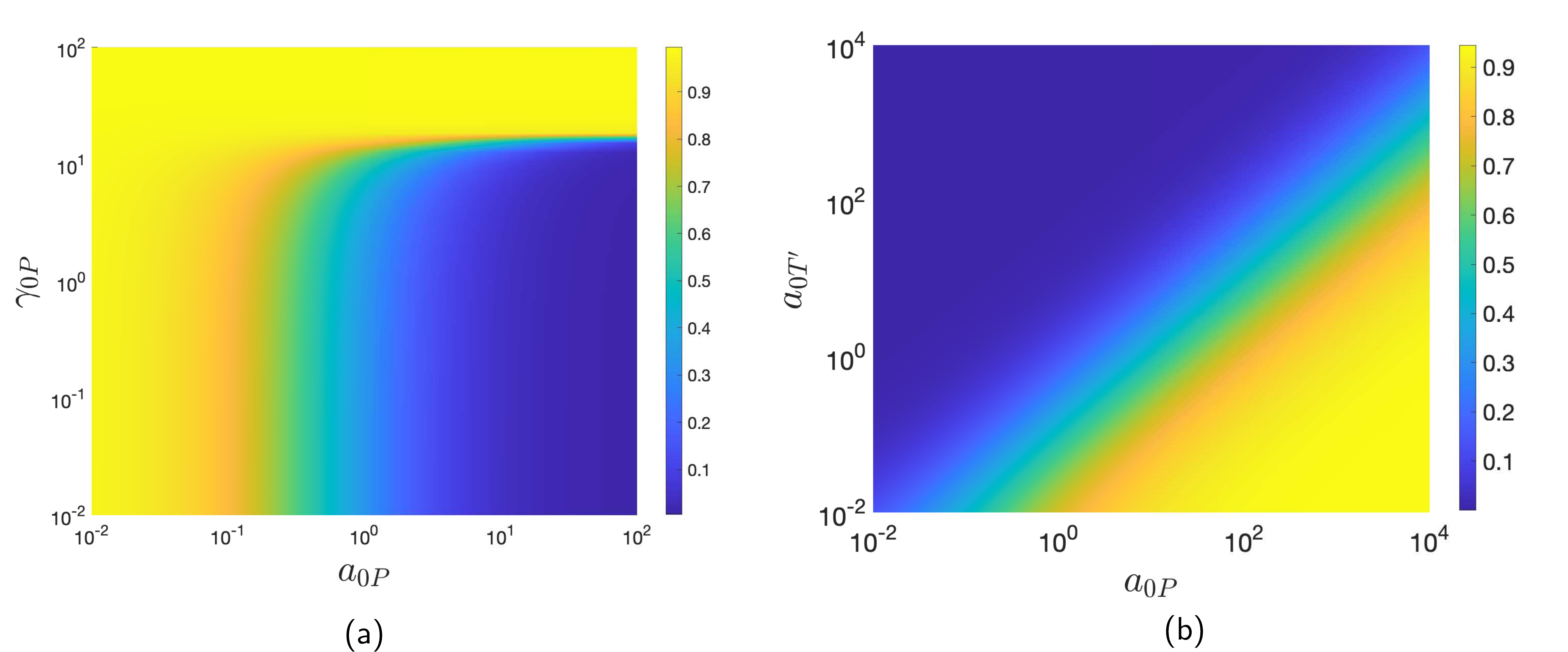}
		\caption{\textbf{ The Global context can reverse   a perturbation}. (a) The fraction of the gene  that is in a \textit{repressed} state. It considers a scenario of two EFs with the activator knocked out. (b) The fraction of the gene   that is in an \textit{active} state. It considers a scenario of two activating EFs and one repressing EF with the first activator knocked out. In both examples, the local parameters are fixed.}
		\label{fig:casessimulation}
	\end{figure}
	\paragraph{A single gene and three EFs}
	By adding a second EF (say an activator), we modify the network \eqref{e.2EF1}-\eqref{e.2EF0} to become as follows:
	
	\begin{align}  
		{\tiny\begin{array}{c}\mbox{Target}\\\mbox{Gene}\end{array}} &\left \{\begin{array}{rl}\mr T+\mr G_{0}\s1 \hspace{-0.1in} & \rightleftharpoons  \mr G_{T}\s1 \lra  \mr G_{+}\s1+ \mr  T, ~ \mr G_{+}\s1  \lra \mr G_{10}\s1 \\ \mr  T'+\mr G_{0}\s1 \hspace{-0.1in} & \rightleftharpoons  \mr G_{T'}\s1 \lra  \mr G_{+}\s1+\mr  T'  \\
			\mr  P+\mr G_{0}\s1  \hspace{-0.1in} & \rightleftharpoons  \mr G_{P}\s1 \lra  \mr G_{-}\s1+ \mr  P, ~ \mr G_{-}\s1  \lra  \mr G_{0}\s1 \end{array} \right . \\
		{\tiny\begin{array}{c}\mbox{Rest of}\\   \mbox{the Genome}\end{array}}&\left\{\begin{array}{rl}\mr  T+\mr G_{0}\s0 \hspace{-0.1in} & \rightleftharpoons  \mr G_{T}\s0 \lra  \mr G_{+}\s0+\mr  T, ~ \mr G_{+}\s0  \lra  \mr G_{0}\s0 \\
			\mr  T'+\mr G_{0}\s0 \hspace{-0.1in} & \rightleftharpoons  \mr G_{T'}\s0 \lra  \mr G_{+}\s0+\mr  T' \\
			\mr  P+\mr G_{0}\s0  \hspace{-0.1in} & \rightleftharpoons  \mr G_{P}\s0 \lra  \mr G_{-}\s0+\mr  P, ~ \mr G_{-}\s0  \lra  \mr G_{0}\s0  \end{array} \right.. 
	\end{align}

	Similar to \eqref{ActiveE}, the activation function takes the following form:
	\begin{equation}\label{ActiveE3} \Psi=    G_{tot}\s1\frac{\rho c_0+ c_T T + c_{T'} T'  }{c_0+ c_T T +  c_{T'} T'+ c_P P } ,
	\end{equation} 
	
	Supp. Fig. \ref{fig:casessimulation} shows how the global context can dictate the effect of a knockout.  Supp. Fig. \ref{fig:casessimulation}-a demonstrates that a low global marking ratio $\gamma_{0P}$ and a high global association ration $a_{0P}$ dilutes the repressing EF P in the absence of the activator $T$. Supp. Fig. \ref{fig:casessimulation}-b also considers the case of activator knockout, and it shows that an alternative activator with a low global association ratio can rescue activation if the global association ratio of the repressor is high.
	
	\subsection{Modeling multiple knockout experiments}
	In the main text, we have shown pictorially how the results of different knockout experiments can be explained by the global context (Figures 2 and 3).  {{  To illustrate the underlying effect further, we show in Supplementary Figure \ref{f.distributions}  the distributions of PRC2, KMT2D and KMT2C for each of the experimental scenarios. In the case of PRC2-KO,  the rest of the genome  sequesters most of the free KMT2C/D which results in their dilution at the promoter of the gene and lackluster activation. In the KMT2D-KO case, most of the free PRC2 gets sequestered and it outcompetes KMT2C across the genome which leaves a sufficient number of unbound KMT2C complexes to activate the target gene.}

		\begin{figure}[h]
			\centering
			\includegraphics[width=0.4\textwidth]{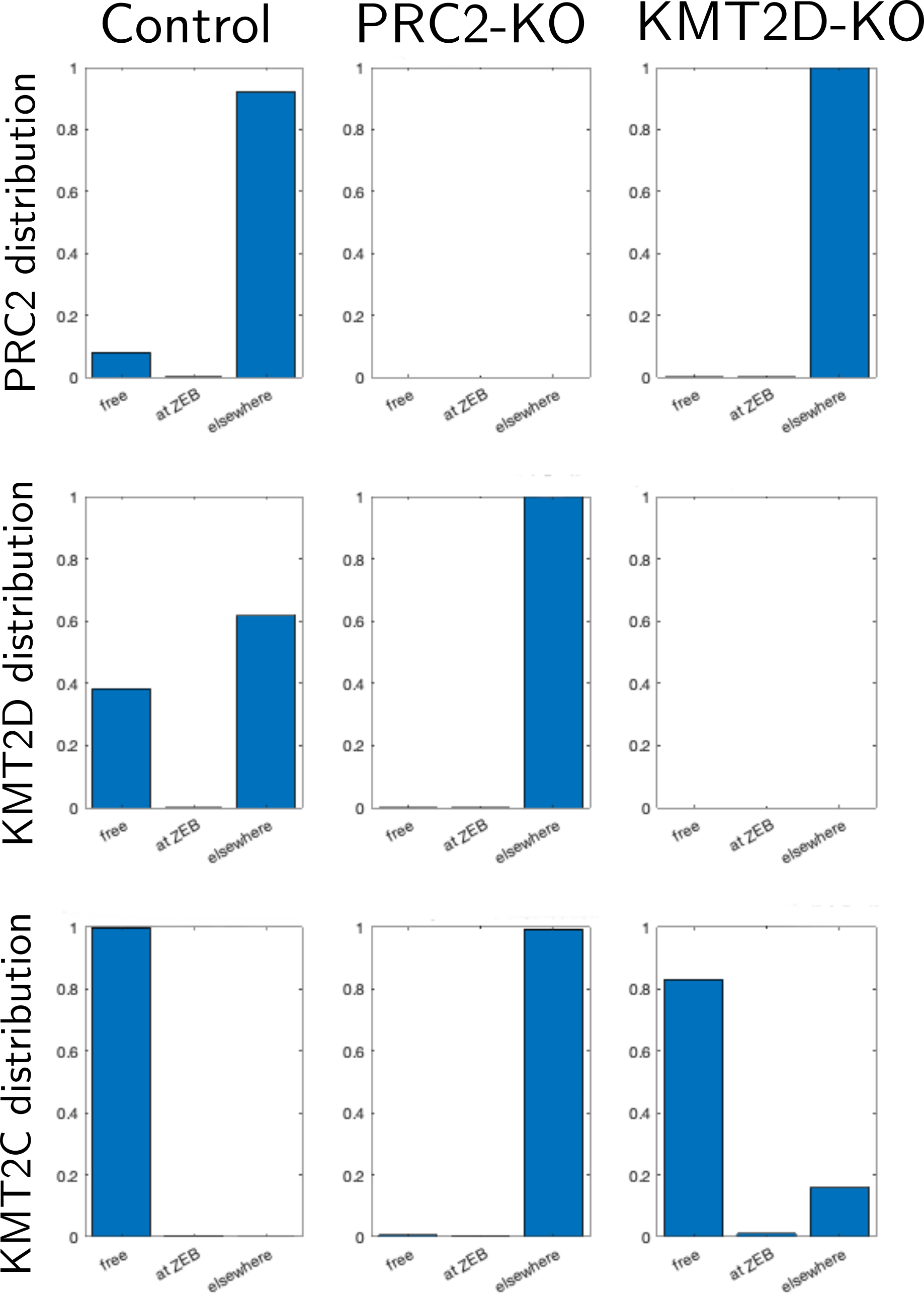}
			\caption{The distribution of the three factors under the three experimental scenarios for a single gene model of ZEB1.  }
			\label{f.distributions}
		\end{figure}
		In addition, Tables 1 and 2 have demonstrated that all possible phenotypes are possible depending on the balance of local and global parameters. It is worth noting that the conditional marginal probability distributions presented in Table 2 depend on the values of the fixed parameters. Here we recompute the conditional marginal probability distributions   by choosing different groups of fixed parameters.

		To be more concrete, recall that we have twelve parameters: The local association ratios $a_{1P},a_{1T},a_{1T'}$, the local marking ratios $\gamma_{1P},\gamma_{1T},\gamma_{1T'}$, the global association ratios $a_{0P},a_{0T},a_{0T'}$, and the global marking ratios $\gamma_{0P},\gamma_{0T},\gamma_{0T'}$.  Table 2, in the main text, had the marking ratios fixed, leaving the association ratio  variable. 
		
		First, we fix the six local parameters with P being dominant, while leaving the six global parameters variable. Supp. Table \ref{fig:paramaters} shows the marginal distribution of each global parameter conditioned realizing on a specified phenotypes. For example, consider the plot depicted in the fifth column and first row in the table. Among the parameter sets that give the corresponding phenotype, the range of values of the global association ratio $a_{0P}$ is higher than $10^2$ (compared to a local association ratio of 2). This means that the repressing EF P has to have a strong tendency to bind to targets across the genome. The figure shows, in addition, that the most ``natural'' phenotype (shown in the third column) does not require specific values of the parameters and it has  wide distributions. While the more ``paradoxical'' phenotypes are only realized under the conjunction of multiple stringent parameter ranges.
		
		Two other Tables are provided in Supp. Table \ref{fig:paramaters2} and \ref{fig:paramaters3}. For the first, we fix the global parameters while varying the local parameters. For the second, we make the two activators identical locally, and have the same total levels globally. In both tables, we see that all phenotypes are achievable.
		
		\begin{table}
			\centering
			\includegraphics[width=1\linewidth]{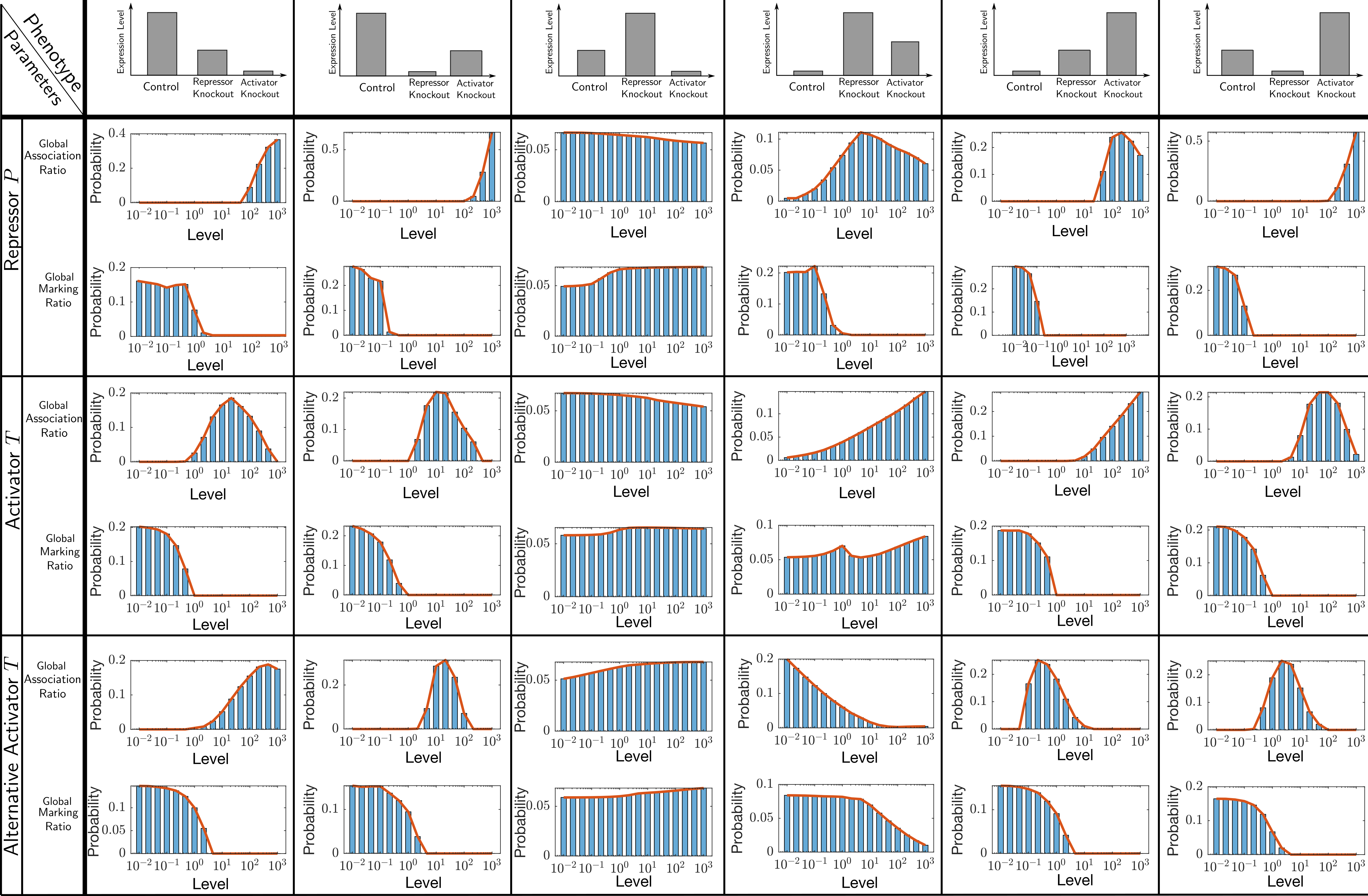}
			\caption{\textbf{All  phenotypes  are possible under the appropriate global context.} Each plot depicts the marginal probability distribution of the parameter under consideration conditioned on the phenotype under consideration.  The local parameters are fixed so that the repressing EF P is dominant. In particular, we have $a_{1P}=2, a_{1T}=0.2, a_{1T'}=0.2, \gamma_{1P}=\gamma_{1T}=\gamma_{1T'}=1$. The six global parameters are varied with 16 levels between $10^{-3}$ and $10^3$. This provides $6^{16}\approx 2.8211\times 10^{12}$ sets of parameters. For each of which, the steady states are calculated numerically for the control and the knockout cases. A set of parameters is said to give one of the six phenotypes    if the highest expression level (amongst the three cases) is at least 50\% higher than the second highest, and latter is at least 50\%  higher than the third highest.    }
			\label{fig:paramaters}
		\end{table}
		\begin{table}
			\centering
			\includegraphics[width=1\linewidth]{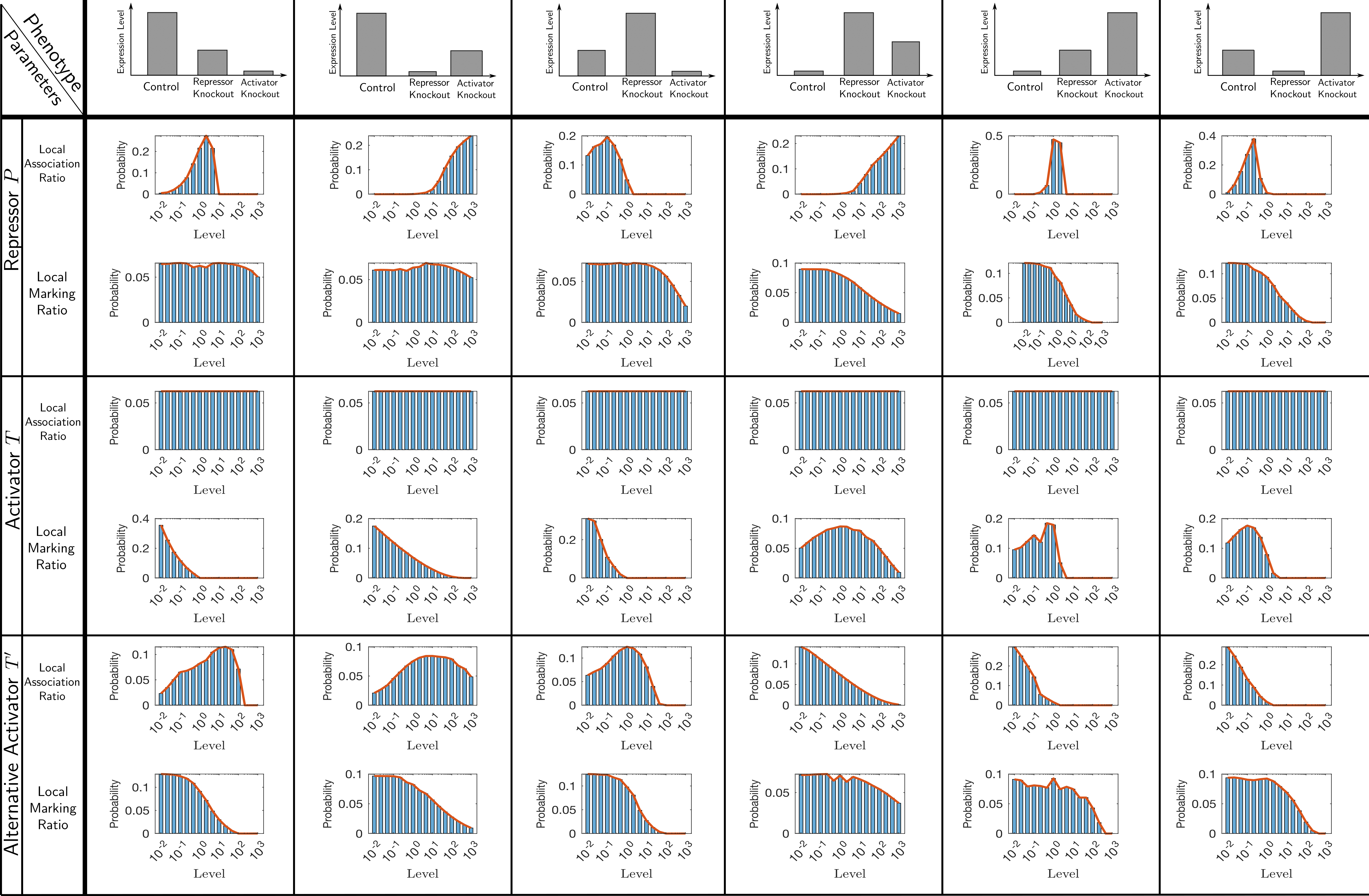}
			\caption{\textbf{All  phenotypes  are possible for a fixed global context.} Each plot depicts the marginal probability distribution of the parameter under consideration conditioned on the phenotype under consideration.  The global parameters are fixed as follows: $a_{0P}=a_{0T}=100, a_{0T'}=1,\gamma_{0P}=\gamma_{0T}=\gamma_{0T'}=0$.  The total levels are $P_{tot}=1000, T_{tot}=500, T_{tot}^{'}=100$. The six other parameters are varied with 16 levels between $10^{-3}$ and $10^3$. This provides $6^{16}\approx 2.8211\times 10^{12}$ sets of parameters. For each of which, the steady states are calculated numerically for the control and the knockout cases. A set of parameters is said to give one of the six phenotypes    if the highest expression level (amongst the three cases) is at least 50\% higher than the second highest, and latter is at least 50\%  higher than the third highest.    }
			\label{fig:paramaters2}
		\end{table}
		
		\begin{table}
			\centering
			\includegraphics[width=1\linewidth]{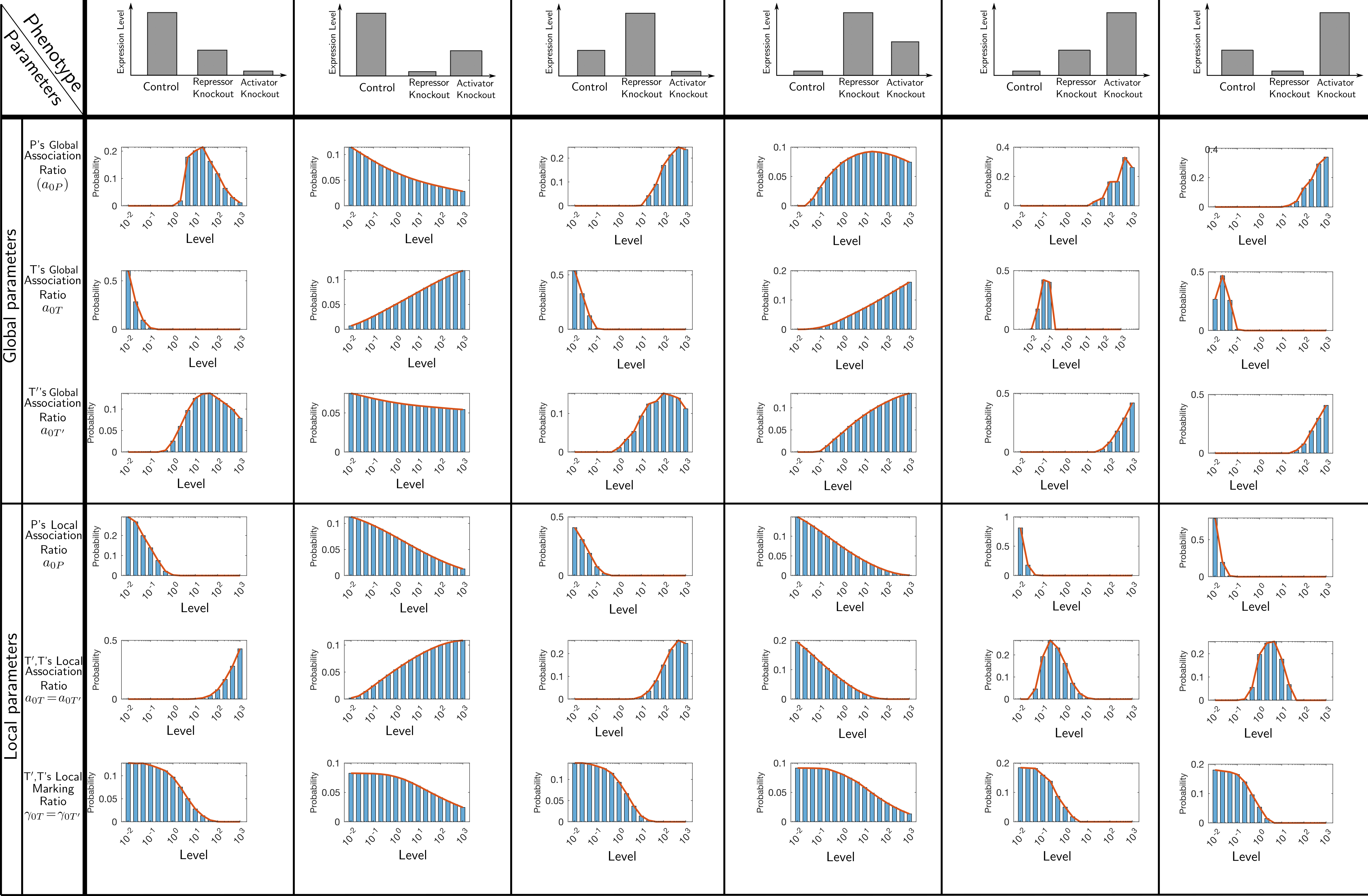}
			\caption{\textbf{All  phenotypes  are possible even when the two activators act the same locally.} Each plot depicts the marginal probability distribution of the parameter under consideration conditioned on the phenotype under consideration.  The following parameters are fixed: $\gamma_{0P}=\gamma_{0T}=\gamma_{0T'}=0, \gamma_{1P}=1$. Furthermore, the activators are assumed to act the same locally. Hence, we have $a_{1T}=a_{1T'}, \gamma_{1T}=\gamma_{1T'}$.  The total levels are $P_{tot}=1000, T_{tot}=T_{tot}^{'}=500$. The six other parameters are varied with 16 levels between $10^{-3}$ and $10^3$. This provides $6^{16}\approx 2.8211\times 10^{12}$ sets of parameters. For each of which, the steady states are calculated numerically for the control and the knockout cases. A set of parameters is said to give one of the six phenotypes    if the highest expression level (amongst the three cases) is at least 50\% higher than the second highest, and latter is at least 50\%  higher than the third highest.    }
			\label{fig:paramaters3}
		\end{table}
	}
	\newpage
	\section{Modeling a self-activating gene subject to epigenetic factor competition}
	
	\subsection{Modeling a self-activating gene}\label{s.selfactivating}
	We review here the standard framework for modeling self-activation. Consider a gene G expressing a protein X that activates its own gene. Then, X binds to G and influences its expression. This can be represented using the following reactions:
	\[p\mr X+ \mr G_0 \xrightleftharpoons[a_-]{a_+/p} \mr G_1,~ \mr G_0 \lra^{k_+} \mr G_0 + \mr X,~  \mr G_1 \lra^{\lambda k_+} \mr G_1 + \mr X, ~ \mr X \lra^{k_-} \emptyset, \]
	where $p$ denotes the cooperativity index, $\mr G_0, \mr G_1$ denotes the unbound and bound states, respectively. Expression of the gene G has a basal rate $k_+$ that is affected by the binding of the X by a factor of $\lambda$. We must have $\lambda>1$ for X to be self-activating.
	
	Solving   \eqref{e.ss} for the promoter dynamics, and letting $X_0:=p a_-/a_+$, we get the standard equations \cite{lu13b,racipe}:
	\begin{equation}\dot X = {k_+} \hill(X;\lambda,n,X_{0}) - k_- X,\end{equation}   {where} \begin{equation} \label{H} \hill(Y;p,\lambda,Y_0):=\frac{ 1+\lambda (Y/Y_{0})^p  }{  1+(Y/Y_{0})^p  }. \end{equation}
	The function $\hill$ is the shifted Hill function. 
	
	\subsection{The interaction between  self-activating genes and epigenetic factor competition}
\begin{figure}
		\centering
		\includegraphics[width=0.6\textwidth]{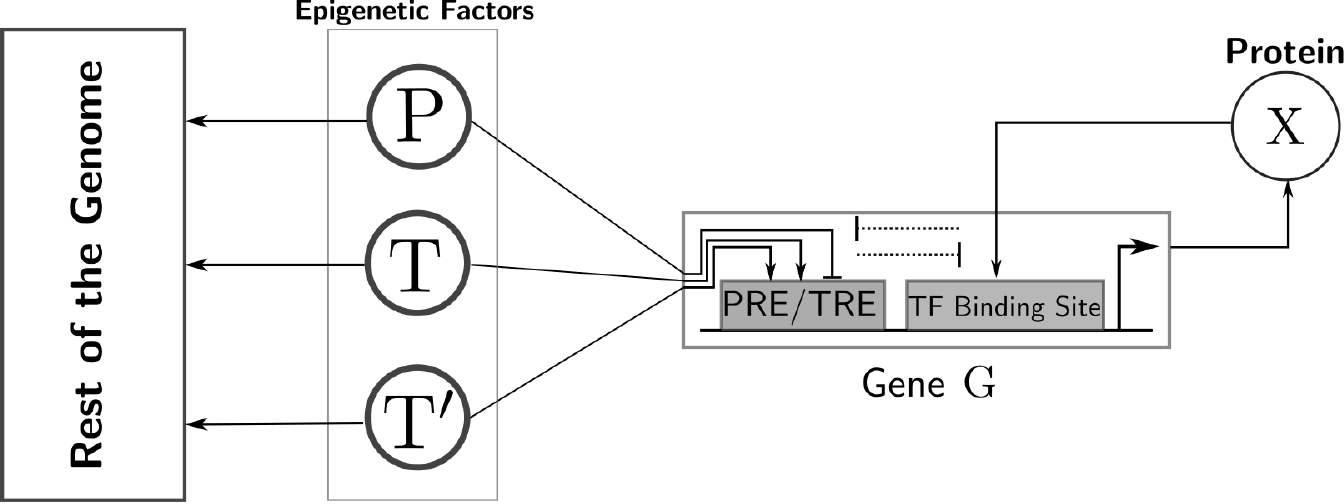}
		\caption{A single self-activating gene subject to one repressing and two activating EFs.}
		\label{f.TF_EF}
	\end{figure}

	To simplify the notation, we study the case of two activating EFs $\mr T,\mr T'$ and one repressing EF $\mr P$ acting on $n$ self-activating genes $\mr G\s1,..,\mr G\s n$ expressing protein $\mr X_1,..,\mr X_n$. The rest of the genome denoted by $\mr G\s0$ as before.
	In order to combine the TF and EF effects, we refer to the pictorial representation in Figure \ref{f.TF_EF}.	
	For a given gene G, we use the notation $\mr G_{ij}$. The first subscript   denotes the histone state, while the second subscript denotes the occupancy of the TF binding site. Based on the evidence reviewed in the main text, the antagonism between active transcription and histone silencers (such as PRC2) is formalized by assuming that the  state in which   the self-activating TF X and the repressing EF P are both bound is rare and hence it is neglected in the model.   Similarly, we assume that a gene marked with repressing histone mark is inaccessible to the TF. 
	\begin{table}
		\centering
		\begin{tabular}{|c|c|c|c|}
			\hline
			TF site                                         & PcG/TrX response element                       & Histone mark                                   & Symbol                                                  \\ \hline\hline
			0                                               & 0                                              &     0                                           & $\mr G_{00}$                                              \\ \hline
			
			0                                               & $\mr P$                                                &    $-$ \                                    & $\mr G_{P0}$                                              \\ \hline
			0                                               & $\mr T$                                               &     $+$                                               & $\mr G_{T0}$                                              \\ \hline
			0                                               & $\mr T'$                                          &    $+$                                          & $\mr G_{T'0}$                                            \\ \hline
			0                                               & 0                                              & +                                              & $\mr G_{+0}$                                              \\ \hline
			
			0                                               & 0                                              & $-$                                            & $\mr G_{-0}$                                              \\ \hline \hline
			
			$\mr X_i $                                             & 0                                              &                                                & $\mr G_{0X_i}$                                              \\ \hline
			
			$\mr X_i $                                                & $\mr T$                                              &       $+$                                              & $\mr G_{TX_i}$                                              \\ \hline
			
			$\mr X_i $                                               & $\mr T'$                                          &       $+$                                    & $\mr G_{T'X_i}$                                            \\ \hline
			
			{ $\mr X_i$} & { 0} & { +} & { $\mr G_{+X_i}$} \\ \hline
		\end{tabular}
		\caption{All possible states of a self-activating  gene subject to epigenetic competition between three EFs. $\mr P$ denotes the repressing EF, while $\mr T,\mr T'$ denote the activating EFs.}\label{table}
	\end{table} 
	The gene states are listed in Table \ref{table}.
	Therefore, we can write the following    reactions for a repressing EF $\mr P$:
	\begin{align}\label{e.selfP}
		\mr P+\mr G_{00}\s i  \xrightleftharpoons[]{} \mr G_{P0}\s i \lra^{ } \mr G_{-0}\s i+\mr P, ~ \mr G_{-0}\s i  \lra^{} \mr G_{00}\s i. 	
	\end{align}
	Similarly, we write the following for the activating EFs $\mr T, \mr T'$:
	\begin{align}\nonumber
		{\tiny\begin{array}{c}\mbox{First}\\\mbox{Activator}\end{array}} &\left \{\begin{array}{rl}
			\mr T+\mr G_{00}\s i & \xrightleftharpoons[]{} \mr G_{T0}\s i \lra^{} \mr G_{+0}\s i+\mr T, ~ \mr G_{+0}\s i  \lra^{} \mr G_{00}\s i\\ 
			\mr T+\mr G_{0X_i}\s i & \xrightleftharpoons[]{} \mr G_{TX_i}\s i \lra^{ } \mr G_{+X_i}\s i+\mr T, ~ \mr G_{+X_i}\s i  \lra^{} \mr G_{0X_i}\s i	\end{array} \right . \\ \label{crn_TFEF}
		{\tiny\begin{array}{c}\mbox{Second }\\\mbox{Activator}\end{array}} &\left \{\begin{array}{rl}	
			\mr T'+\mr G_{00}\s i & \xrightleftharpoons[]{} \mr G_{T'0}\s i \lra^{ } \mr G_{+0}\s i+\mr T',\\
			\mr T'+\mr G_{0X_i}\s i & \xrightleftharpoons[]{} \mr G_{T'X_i}\s i \lra^{ } \mr G_{+X_i}\s i+\mr T',  \end{array} \right .	\\
		{\tiny\begin{array}{c}\mbox{TF Binding/ }\\\mbox{Unbinding}\end{array}} &\left \{\begin{array}{rl}
			p\mr X_i+ \mr G_{00}\s i & \xrightleftharpoons[]{} \mr G_{0X_i}\s i, ~ p\mr X_i+ \mr G_{T0}\s i   \xrightleftharpoons[]{} \mr G_{TX_i}\s i,  \\
			p\mr X_i+ \mr G_{T'0}\s i  & \xrightleftharpoons[]{} \mr G_{T'X_i}\s i, ~  p\mr X_i+ \mr G_{+0}\s i   \xrightleftharpoons[]{} \mr G_{+X_i}\s i.
		\end{array} \right . \nonumber
	\end{align}
	
	In addition, we have \eqref{e.2EF0} which describes the competition of the EFs for sites on the rest of the genome. %
	
	We are interested in characterizing the effective activation function of the gene $\Psi(X)$, i.e an analogous function to \eqref{H}. Therefore, we need to describe the gene expression network:
	\begin{align} \nonumber
		\mr G_{TX_i}\s i &\lra^{k_+} \mr G_{TX_i}\s i+\mr X_i,~\mr G_{T'X_i}\s i  \lra^{k_+} \mr G_{TX_i}\s i+\mr X_i,~\mr G_{+X_i}\s i  \lra^{k_+} \mr G_{+X_i}\s i+\mr X_i, \\ \label{e.selfexpression} \mr G_{T0}\s i &\lra^{\rho_X k_+} \mr G_{T0}\s i+\mr X_i,~\mr G_{T'0}\s i  \lra^{\rho_X k_+} \mr G_{T0}\s i+\mr X_i,~\mr G_{+0}\s i  \lra^{\rho_X k_+} \mr G_{+0}\s i+\mr X_i, \\    \nonumber
		\mr G_{0X_i}\s i &\lra^{\rho_U k_+} \mr G_{0X_i}\s i+\mr X_i,~\mr G_{00}\s i  \xrightarrow{\rho_X \rho_U k_+} \mr G_{00}\s i+\mr X_i,  ~
		\mr X_i   \lra^{k_-}  \emptyset,
	\end{align}
	where $0<\rho_X,\rho_U<1$ are the ratios describing the reduction in expression rate due to the absence of the self-activating TF and the absence of the activating EF, respectively.
	
	Therefore, the $i$th activation function   can be written as:
	\begin{equation}\label{e.activationfunction} \Psi_i(X_1,..,X_n)=    G_{TX_i}\s i +   G_{T'X_i}\s i +   G_{+X_i}\s i + \rho_X ( G_{T0}\s i +   G_{T'0}\s i +   G_{+0}\s i) + \rho_U (G_{0X_i}\s i+\rho_X G_{00}\s i) .\end{equation}
	
	Compared to \eqref{H}, the activation function depends on the local parameters, the EF levels, the other genes in the network, as well as the global parameters. The cross-talk between the different genes is indirect and is mediated via the EFs. As explained earlier, an analytical solution is not feasible. Hence, we solve the equations \eqref{polynomial} numerically for each given tuple $(X_1,..,X_n)$.

	\begin{figure}
		\centering
		\subfigure[There are no global targets]{\includegraphics[height=2.25in,width=0.41\linewidth]{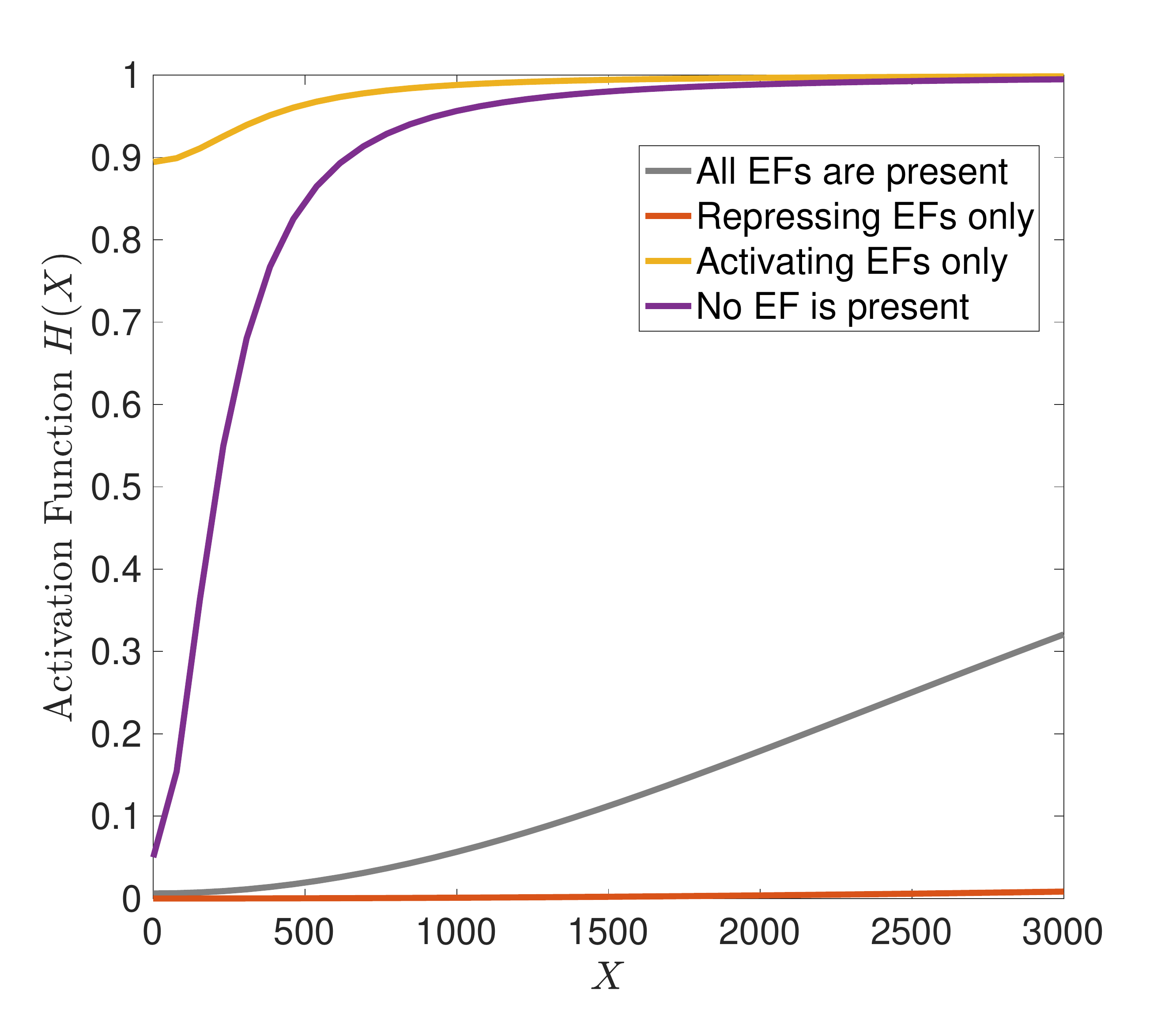}}
		\subfigure[There are  global targets]{\includegraphics[height=2.25in,width=0.41\linewidth]{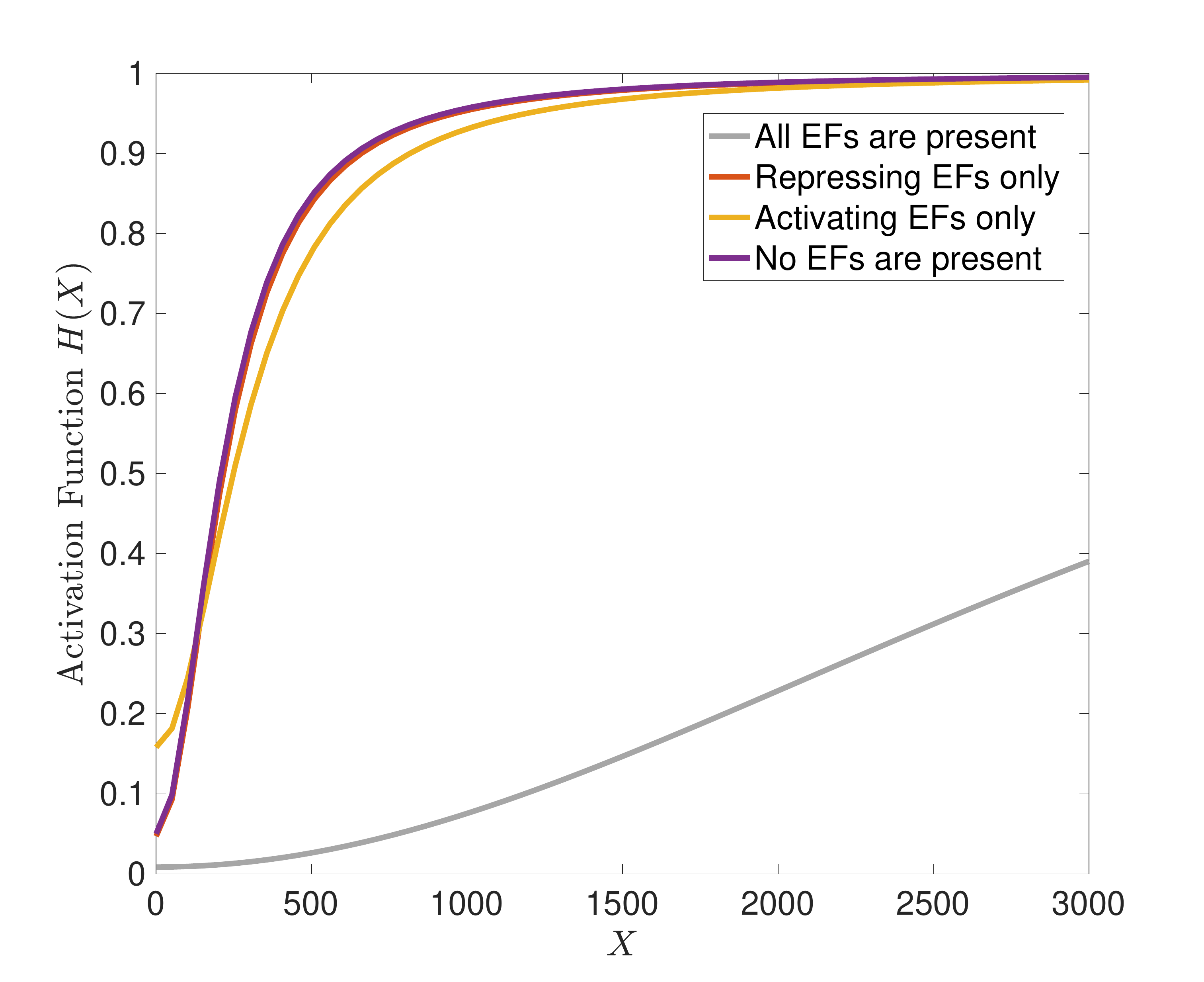}}
		\caption{\textbf{The global context dictates the local activation curve of the gene under different perturbations.} (a) The activation curves with no global context, i.e. $G_{tot}\s0=0$, under different perturbations. (b) The activation curves with a global context, i.e. $G_{tot}\s0\gg G_{tot}\s1$,  under different perturbations.  }
		\vspace{-0.1in}	\label{fig:1geneEF}
	\end{figure}
	
	\paragraph{Example: Multiple Knockouts} We consider here the fifth phenotype depicted in Supp. Table \ref{fig:paramaters} where the repressing P protein is more dominant locally. However, it also has strong affinity across the genome. The highest expression is achieved when the activator in knocked-out, followed by when the repressor is knocked-out. Here, we study how the EF competition model can interact with TF self-activation for a single gene. Figure \ref{fig:1geneEF} shows the effect of the global context by comparing two scenarios. The first is depicted in Figure \ref{fig:1geneEF}-a where the global context is negligible. When there are no EFs present, the resulting curve is the TF Hill activation function \eqref{H}. When all EFs are present, the repressing EF is dominant as assumed, hence a much higher level of the protein X is required to activate the gene. However, when the activators are knocked-out, the repressing EF does not allow any expression for the simulated range of the protein. On the other hand, when there is no repressor, the gene is activated even with very little TF available. The picture changes drastically when the global context is considered as in Figure \ref{fig:1geneEF}-b. It can be seen that knocking out the repressor will dilute the activators making the activation function resemble the a gene operating in the absence of EFs. We get a similar effect when both activators are knocked out, since the repressor gets diluted as a result.

	\newpage
	
	\section{Modeling a generic GRN subject to epigenetic factor competition}

	\subsection{Review of GRN modeling}

	In order to combine our epigenetic regulatory model with standard GRN models, we review   GRN models   using BIN formalism. A self-activating gene has been reviewed in \S \ref{s.selfactivating}. In particular, we have defined the Hill function \eqref{H} which is activating if $\lambda>1$ and repressing if $\lambda<1$.

	\subsubsection{Multiple TFs} \label{s.multiple}
	Assume that a gene G is subject to the action of multiple TFs $\mr Y_1,..,\mr Y_n$ which can bind simultaneously.  Hence, there are $2^n$ binding states based on the TFs that are bound. In order to simplify the analysis, we assume that the TFs bind to their respective sites and affect expression of the gene \textit{independently} of the other TFs. We will make this precise below.
	
	The binding state of a gene denoted as $\mr G_s$ where , $s \in \{0,1\}^n$. For instance, $\mr G_s$ with $s=[0,1,1]$ means that the first TF is unbound, while the second and third TFs are bound. We also write it as $s=011$ for brevity. Let $\mathcal G_{+j}$ be the set of states where the $k$th TF is bound, i.e., $\mathcal G_{+j}:=\{\mr G_s| s_j=1\}$. Similarly, $\mathcal G_{-j}:=\{\mr G_s| s_j=0\}$. For example, in the case of three TFs, $\mathcal G_{+2}=\{\mr G_{010},\mr G_{110},\mr G_{011},\mr G_{111}\}$.
	
	Given the notation above, we can state our exact assumptions regarding the binding of the TFs to the gene:  
	\begin{enumerate}
		\item \emph{Independence:} The TFs bind independently of each other, meaning  that the cooperativity index $p_j$, and the binding/unbinding rates $ a_{+j}, a_{-j}, j=1,..,n$,   of each TF are independent of the presence of other TFs. More precisely, fix $j$. Then, for all $\mr G_{s^{-}} \in \mathcal G_{-j}$, the model contains the following binding/unbinding reaction:
		\[ p_j \mr Y_j + \mr G_{s^{-}} \xrightleftharpoons[a_{-j}]{a_{+j}} \mr G_{s^{+}}, \]
		where $s^{+}_j=1$, i.e., $\mr G_{s^{+}} \in \mathcal G_{+j}$. 
		\item \emph{Uniformity:} The effect of each TF on the expression of the gene is independent of the presence of other TFs. In other words, each TF $\mr Y_j$ has a corresponding factor $\lambda_j$.  Furthermore, for any two $s^-,s^+$ which are identical expect for the $j$th site, the model contains the following two reactions
		\[\mr G_{s^-} \lra^{  \tilde k} \mr G_{s^-} + \mr X, ~ \mr G_{s^+} \lra^{\lambda_j \tilde k} \mr G_{s^+} + \mr X.  \]
		Hence, the ratio between the production rate of  any state with $\mr Y_j$ unbound and the corresponding state with $\mr Y_j$ is bound is always equal to $\lambda_j$ regardless of the presence of other TFs
	\end{enumerate}
	
	If the above assumptions are satisfied, then  it can be shown that the standard expression holds \cite{racipe}:
	
	\begin{equation}\dot X = k_{+} \prod_{j=1}^n   \hill(Y_j;\lambda_j,p_j,Y_{0j}) - k_- X,   \end{equation}
	where $\hill$ is defined in \eqref{H}, and $Y_{0j}=p_ja_{-j}/a_{+j}$. 
	
	\subsubsection{Multiple TFs and multiple genes} The above formalism can be extended to any GRN with $n$ genes. Hence, a general model can be written as follows:
	\begin{equation}\label{grn}\dot X_i = k_{+i} \prod_{j=1}^n   \hill(X_{ij};\lambda_{ij},p_{ij},X_{0ij})  - k_{-i} X_i, i=1,..,n. \end{equation}
	where $X_{0ji}=\infty$ if $X_j$ does not act on the gene expressing $X_i$. External factors can be added to \eqref{grn} if needed.

	\subsubsection{Micro-RNA regulation} Another form of regulation is via micro-RNAs (miRNAs). They inhibit target genes by binding to mRNAs and impeding translation. \cite{lu13b,lu13,MA_springer22}. We review the modeling framework presented in \cite{lu13b,lu13}. Consider a gene G that expresses a protein $X$ via an mRNA molecule which has has $m$ binding sites that miRNAs can bind to. Let $i \in{0,1,..,m}$, then $\mr M_i$ denotes the mRNA molecule with exactly $i$ miRNA molecules bound to it. Denote the miRNA molecule by \textmu. We assume that the rates of binding and unbinding of \textmu ~to a particular binding site on the mRNA molecule are given as $\nu_+,\nu_-$, respectively. Consider $\mr M_i$. Then, it has $m-i$ sites available for miRNAs. Hence, the total binding rate is $(m-i)r_+$. Similarly, $\mr M_{i+1}$ has $i+1$ bound miRNAs and the total unbinding rate is $(i+1) r_-$. Therefore, we write the following reactions: 
	\begin{align*} \mr G & \lra^{k_+} \mr G+ \mr M_0, ~ \mr M_0 \lra^{\beta_0} \emptyset,~  \emptyset \xrightleftharpoons[k_{-\mu}]{k_{+\mu}} \mbox{\textmu} , \\
		\mbox{\textmu} + \mr M_{i-1}& \xrightleftharpoons[(i+1)r_-]{(m-i)r_+} \mr M_i  \xrightarrow{ \beta_i\s m } \mbox{\textmu}, ~ \mr M_i  \lra^{\beta_i\s\mu } \mr M_{i-1}  \\
		\mr M_i &\lra^{\ell_i} \mr M_i + \mr X, ~ \mr X \lra^{k_-} \emptyset, i=1,..,m,
	\end{align*}
	where %
	the superscripts $(m),(\mu)$ are used to distinguish the decay rates of mRNA and  microRNA in the corresponding mRNA-microRNA complex. 
	It is generally assumed that $0<\beta_0\s m<..<\beta_m\s m$, and $\ell_0>\ell_1>..>\ell_m>0$. 
	
	As assumed in \cite{lu13b,lu13}, we assume that the binding and binding rates $r_+,r_-$ are very fast. Hence, we write $r_+=\tilde r_+/\varepsilon, r_-=\tilde r_-/\varepsilon $ for some small $\varepsilon>0$.
	Let ${\mathbf M}=[M_0,..,M_m]^T$ be the vector of all mRNA configurations. Therefore, we can write the following ODE to describe the time-evolution of $M(t)$:
	\begin{align*}\dot {\mathbf M}(t)& = \frac 1{\varepsilon} A{\mathbf M}-  B{\mathbf M}  +c , ~  \end{align*}
	\mbox{where}   \\
	\scalebox{0.735}{\begin{minipage}[c]{\linewidth}\[A= { 
				\begin{bmatrix} -m\mu \tilde r_+ & \tilde r_- & 0 & \dots &&&&& 0 \\
					\vdots &  &\ddots  &&&&&& \vdots \\
					0 & \dots & 0 & \mu \tilde r_+ (\!m\!-i\!+\!1) & -i \tilde r_-\! - \mu \tilde r_+(\!m\!-\!i\!) & ( i\!+\!1) \tilde r_- & 0 & \dots & 0 \\
					\vdots &&&&&&&& \vdots \\
					0 & \dots & &  & & & & \mu \tilde r_+ & -m \tilde r_-  \end{bmatrix}, B= \begin{bmatrix} \beta_0&    & ... & 0 \\ 0 & \beta_1 & \\ \vdots &  & \ddots & \vdots \\
					0 &  \dots &   & \beta_{n} \end{bmatrix}, c= \begin{bmatrix} k_+ G \\ 0 \\ \vdots \\ 0 \end{bmatrix} }. \] \end{minipage} } \\
	
	\noindent We can write $\varepsilon \dot {\mathbf M} = A{\mathbf M} + \varepsilon (c-B{\mathbf M})  $. By letting $\varepsilon \to 0$, we get $A{\mathbf M}=0$. Since $\mathbf 1^T A=0$, we have the conservation law $ \mathbf 1^T {\mathbf M}=M$, where $M$ is   \textit{the total mRNA} which is constant in the fast-time scale. Hence, we need to solve the linear system $A{\mathbf M}=0, \mathbf 1^T {\mathbf M} = M$. By algebraic manipulations and defining $\mu_0:=r_-/r_+$, the result can be shown to be:
	\[ M_i=M  {n \choose i } \frac{ (\mu/\mu_0)^i } {(1+ \mu/\mu_0 )^m}, i=0,..,M.\]

	Let $\ell=[\ell_0,...,\ell_n]^T$. Therefore, we can write the following system of equations:
	\begin{align*}
		\dot M&=\! \mathbf 1^T (c-B{\mathbf M})\!= k_+ G -  \frac{ \sum_{i=0}^m \beta_i\s m  {m \choose i} (\mu/\mu_0)^i M}{(1+\mu/\mu_0)^m}-\beta M =: k_+ G - Y\s m(\mu) M, \\
		\dot X & =  	 \ell^T {\mathbf M} - k_- X =  \frac{ \sum_{i=0}^m \ell_i  {m \choose i} (\mu/\mu_0)^i M}{(1+\mu/\mu_0)^m} -k_- x := L(\mu) M - k_- X,\\
		\dot \mu &  = k_{+\mu}   - k_- \mu - \frac{ \sum_{i=1}^m \beta_i\s \mu  {m \choose i} (\mu/\mu_0)^i M}{(1+\mu/\mu_0)^m} =: k_{+\mu}   - k_- \mu - Y\s\mu(\mu) M.
	\end{align*}
	Therefore, solving for $X$ at steady state, we get $X=(k_+/k_-) G \Lambda(\mu) $, where $\Lambda$ is given as:
	\begin{equation}\label{Lambda} \Lambda(\mu):= \frac{L(\mu)}{Y\s m(\mu)}= \frac{\sum_{i=0}^m \tilde\ell_i (\mu/\mu_0)^i}{\sum_{i=0}^m \tilde\beta_i (\mu/\mu_0)^i},\end{equation}
	where $\tilde \ell_i ={m \choose i} \ell_i, \tilde \beta_i ={m \choose i} \beta_i$. Compared to \eqref{H} which has three degrees of freedom only, we notice that miRNA regulation gives rise to a rational function that is a ratio of two generic polynomials. For future reference, we define:
	\begin{align}\label{e.LY}
		\mathcal L(\mu;\mu_0,\ell_0,...,\ell_m):=\frac{\sum_{i=0}^m \ell_i (\mu/\mu_0)^i }{(1+\mu/\mu_0)^m}, ~ \mathcal Y(\mu;\mu_0,\beta_0,...,\beta_m):=\frac{\sum_{i=0}^m \beta_i (\mu/\mu_0)^i }{(1+\mu/\mu_0)^m}
	\end{align}
	
	\subsubsection{Combining miRNA and TF regulations} A  gene can be subject to both TF regulation and miRNA regulation. Let us consider genes $\mr G_1,..,\mr G_{N}$. Assume the first $n_p\le n$ genes   express proteins $\mr X_1,..,\mr X_{n_p}$, while the remaining genes express microRNAs \textmu$_{n_p+1}$,..,\textmu$_{n}$. The expressed proteins can act as TFs to activate or inhibit other genes including the genes expressing the microRNAs.  For simplicity, we assume that a protein-expressing gene \textit{cannot} be inhibited by more than one micro-RNA, but a single micro-RNA can target multiple genes.
	
	For a protein $\mr X_i$ that is regulated by a microRNA \textmu$_\ell$.  The combined model can be written as follows, 
	\begin{align*}
		\dot M_{i}&= k_+ \prod_{j=1}^{n_p} \hill_{ij}(X_j;\lambda_{ij},p_{ij},X_{0ij}) - Y_{i}\s m(\mu_{\ell}) M_i , \\
		\dot X_i & =  	   L_{ i}(\mu_{\ell}) M_i - k_- X, i=1,..,n,
	\end{align*}
	while for the microRNA \textmu$_\ell$, we write:
	\begin{align*}
		\dot \mu_\ell & = k_{+\ell} \prod_{j=1}^{n_p} \hill(X_j;\lambda_{  \ell j},n_{\ell j},X_{0\ell j})   - k_{-\ell} \mu_\ell - \sum_{\{i:\mbox{\footnotesize\textmu}_\ell \dashv X_i\}}  Y_i\s\mu(\mu_{\ell}) M_{i}, j=1,..,q.
	\end{align*}
	For a gene $\mr X_i$ that is only regulated by TFs, we write:
	\begin{equation} \dot X_i = k_{+i} \prod_{j=1}^{n_p}   \hill(X_{ij};\lambda_{ij},p_{ij},X_{0ij})  - k_{-i} X_i, i=1,..,N. \end{equation}
	where $X_{0ji}=\infty$ if $X_j$ does not act on the gene expressing $X_i$.

	Using the equations above, the steady state equations describing the the protein $\mr X_i$ which is regulated by microRNA \textmu$_\ell$  can be written as:
	\begin{align}\label{ss.X} 0 & =  	   k_{+i}\frac{L_{ i}(\mu_{\ell})}{Y_{i}\s m(\mu_{\ell})} \prod_{j=1}^{n_p} \hill_{ij}(X_j;\lambda_{ij},p_{ij},X_{0ij})    - k_- X, i=1,..,n, \end{align}
	where $\hill ,\Lambda$ are defined in \eqref{H},\eqref{Lambda}, while the steady equation for a gene regulated by TFs only can be written as:
	\begin{align}\label{ss.X2} 0 & =  	   k_{+i}  \prod_{j=1}^{n_p} \hill_{ij}(X_j;\lambda_{ij},p_{ij},X_{0ij})    - k_- X, i=1,..,n, \end{align}
	
	Finally, the steady state equation for \textmu$_{\ell}$ can be written as:
	\begin{align}\label{ss.mu}
		0 & =  k_{+\ell} \prod_{j=1}^{n_p} \hill(X_j;\lambda_{  \ell j},n_{\ell j},X_{0\ell j})   - k_{-\ell} \mu_\ell - \sum_{\{i:\mbox{\footnotesize\textmu}_\ell \dashv X_i\}}  \frac{Y_i\s\mu(\mu_{\ell})}{Y_i\s m(\mu_{\ell})} \prod_{j=1}^{n_p} \hill_{ij}(X_j;\lambda_{ij},p_{ij},X_{0ij}).
	\end{align}

	\subsection*{Combining local GRN models with EF factor competition}
	
	A general model for a local GRN network that is subject to EF competition can be written using the model components discussed before.

	\subsubsection{Uncoupled transcription and epigenetic competition}
	We first model the case in which the EFs bind to the PRE/TRE components of the gene  independently of the TFs, and vice versa. In addition, we assume that the EFs  exert their effects uniformly regardless of the presence of TFs, and vice versa . This is similar to our earlier assumptions when modeling the action of multiple TFs on a single gene as discussed in \S \ref{s.multiple}.  Therefore, for every protein, the different regulatory terms appear as a product multiplying the production rate. This simplifying assumption is relevant when considering the immediate aftermath of knockout experiments at turned-off genes like ZEB1 and PRRX1 as discussed in the main text.

	In order to make  the discussion more concrete and simplify the notation, we will illustrate our modeling framework by writing the equations for the proposed network in Figure 4-e that have been used to generate the simulation depicted in Figure 4-d. 
	
	\paragraph{The EF competition circuit} Let $\mr{P,T,T'}$ denote PRC2, KMT2D, and the third activator, respectively. %
	Let $\mr{Z,R}$ denote ZEB1 and PRRX1 proteins, respectively. Without loss of explanatory generality, we assume that the kinetic rates  that describe the interactions of $\mr{P,T,T'}$ are identical for both PRRX1 and ZEB1. Hence, we can find $\Psi$ given in \eqref{Psi1} by solving \eqref{polynomial} for a network of two genes and three EFs. Hence, for  constant kinetic rates, $\Psi$ is a function of the total levels $P_{tot},T_{tot}, T'_{tot}$. \textit{As a result, the value of $\Psi$ changes for each knockout.}  
	
	After finding $\Psi$, we can write $\Psi_X= \Psi G_{tot}\s Z/(G_{tot}\s Z+G_{tot}\s R), \Psi_R= \Psi G_{tot}\s R/(G_{tot}\s Z+G_{tot}\s R)$, where $ G_{tot}\s Z, G_{tot}\s R$ are the total copy numbers of the  ZEB1, PRRX1  genes, respectively. 
	
	\paragraph{Local GRN} With reference to Figure 4-e, let $Z, R, S, T$ denote the levels of ZEB1,  PRRX1, SNAI1, TGB-\textbeta. Let $\mu$ be the level of \textmu$_{200}$.  %
	Using \eqref{ss.X},\eqref{ss.X2},\eqref{ss.mu}, we can write the steady state equations as follows:
	\begin{align*}
		0&=k\su{+Z} \Psi\su Z   H\su{Z\to Z}(Z)  H \su{S\to Z}(S) \frac{L\su{Z}\s{\mu}(\mu)}{Y\su{Z}\s m(\mu)} - k\su{-Z} Z, \\
		0&=k\su{+R} \Psi\su R   H\su{R\to R}(R) H\su{S\dashv R}(S) H\su{T\to R}(T)- {k\su{-R}} R , \\
		0 &= I\su{\mathrm{ext}} +k\su{+S}  H\su{R\dashv S}(R) H\su{T\to S}(T) - k\su{-S} S, \\
		0&=  k\su{+T}  \frac{L\su{T}\s{\mu}(\mu)}{Y\su{T}\s m(\mu)}  - {k\su{-T}} T, \\
		0 & =  k\su{+\mu}  H\su{Z\dashv\mu}(Z)  H \su{S\dashv \mu}(S)      - k\su{-\mu} \mu -   \frac{Y\su{Z}\s\mu(\mu)\Psi\su Z   H\su{Z\to Z}(Z)  H \su{S\to Z}(S)}{Y\su{Z}\s m(\mu)}  -  \frac{Y\su{T}\s\mu(\mu)}{Y\su{T}\s m(\mu)}.  
	\end{align*}

	\paragraph{Parameters.}
	We list the parameters used to produce the simulation in Figure 4-d. We start with the competition circuit:
	
	\begin{align} 
		{\tiny\begin{array}{c}\mr G\s{\mr{ZEB1}}\\ +\\\mr G\s{\mr{PRRX1}}\end{array}} &\left \{\begin{array}{rl}\mr T+\mr G_{0}\s1 \hspace{-0.1in} & \xrightleftharpoons[425.8]{115.6}  \mr G_{T}\s1 \xrightarrow{127.34}  \mr G_{+}\s1+ \mr  T, ~ \mr G_{+}\s1  \xrightarrow{815.94} \mr G_{0}\s1 \\ \mr  T'+\mr G_{0}\s1 \hspace{-0.1in} & \xrightleftharpoons[364.7]{939.1} \mr G_{T'}\s1 \xrightarrow{283.1}  \mr G_{+}\s1+\mr  T'  \\
			\mr  P+\mr G_{0}\s1  \hspace{-0.1in} & \xrightleftharpoons[3.28]{9265.8}  \mr G_{P}\s1 \xrightarrow{734.22}  \mr G_{-}\s1+ \mr  P, ~ \mr G_{-}\s1  \xrightarrow{737.56}  \mr G_{0}\s1 \end{array} \right . \\  
		{\tiny\begin{array}{c}\mbox{Rest of}\\   \mbox{the Genome}\end{array}}&\left\{\begin{array}{rl}\mr  T+\mr G_{0}\s0 \hspace{-0.1in} & \xrightleftharpoons[0.1]{1000}  \mr G_{T}\s0 \xrightarrow{0.1}  \mr G_{+}\s0+\mr  T, ~ \mr G_{+}\s0  \xrightarrow{159.9}  \mr G_{0}\s0 \\
			\mr  T'+\mr G_{0}\s0 \hspace{-0.1in} & \xrightleftharpoons[1000]{5.753}  \mr G_{T'}\s0 \xrightarrow{29}  \mr G_{+}\s0+\mr  T' \\
			\mr  P+\mr G_{0}\s0  \hspace{-0.1in} & \xrightleftharpoons[0.1336]{999.98}  \mr G_{P}\s0 \xrightarrow{0.1}  \mr G_{-}\s0+\mr  P, ~ \mr G_{-}\s0  \xrightarrow{556}  \mr G_{0}\s0  \end{array} \right.. 
	\end{align}
	The remaining parameters are: $P_{tot}=894.14,~T_{tot}=494.46,~T_{tot}'=81.34, G_{tot}\s0=975, G_{tot}\s Z=0.5, G_{tot}\s R=0.5,  \rho_U=0.05$.\\
	For the local GRN, the activation functions are given as:
	\begin{align*} 
		H\su{Z\to Z}(Z)&=\hill(Z;Z_0,n\su Z,1.5),  H \su{S\to Z}(S)=\hill(S;500,2,2), Z_0=5148.9, n\su Z=4,\\ H\su{S\dashv R}(S)&=\hill(S;250,4,0.5), H\su{T\to R}(T)=\hill(T;10,4,4), \\ H\su{R\to R}(R)&=\hill(R;R_0,n\su R,1.5),R_0=1040.4,n\su R=2, \\ H\su{R\dashv S}(R)&=\hill(R;485.27,6,0), H\su{T\to S}(T)=\hill(T;6.2059,4,7.3225 )\\
		H\su{Z\dashv\mu}(Z)&=\hill(Z;63.27,1,0), H\su{S\dashv \mu}(S)=\hill(S;500,2,0.5), \\
		Y_Z\s m(\mu)&=\mathcal Y(\mu;20,1,2,2,3,3,3,3), Y_Z\s \mu(\mu)=\mathcal Y(\mu;20,0,0.6038,1.2076,1.8114,7.2456,10.868), \\   	L_Z(\mu)&=\mathcal L(\mu;20,1,0,0,0), Y_T\s m(\mu) =\mathcal Y(\mu;22.84,1,5.139,5.139,49.287,100.66), \\L_T(\mu)&=\mathcal L(\mu;22.84,1,0,0,0), Y_T\s \mu(\mu)=0,
	\end{align*}
	while the remaining parameters are 	$k\su{+Z} =79.461, k\su{-Z}=0.01, k\su{+\mu}=65,k\su{-\mu}=2.5, k\su{+R}=6.3422, k\su{-R}=0.02,  k\su{+S} =18.005 , k\su{-S}=1, k\su{+T}=100, k\su{-T}=1,  I\su{\mathrm{ext}}=0.1$.

	\subsubsection{Coupling transcription and epigenetic competition}
	As reviewed in the main text, PRC2 interacts antagonistically with active transcription. Hence, the model presented in the previous subsection cannot accurately capture a local GRN in which ZEB1 and PRRX1 are highly expressed. This is especially relevant in the case of sequential knockout experiments.
	
	Therefore, we modify the model presented in the previous subsection by the utilizing the  model presented  in \S S2 that couples transcription and epigenetic competition. Here we show that such a model can be integrated into the local GRN model by assuming that the EF competition-self-regulation sub-circuit is independent of other regulators and exerts its effect uniformly (similar to the assumptions made in the previous subsection and in \S\ref{s.multiple}).
	
	In order to make  the discussion more concrete and simplify the notation, we will illustrate our modeling framework by writing the equations for the proposed network in Figure 4-e that have been used to generate the simulation depicted in Figure 5-b. Needless to say, the underlying principles are generalizable to arbitrary networks.

	\paragraph{Epigenetic competition/self-regulation subcircuit}  Similar to the previous subsection, for the purpose of simplifying the computations, we assume that PRRX1 and ZEB1 react with EFs in a similar way except for the manner in which each TF bind to its own promoter. To model the interaction between ZEB1, PRRX and the EFs, we use the model described by \eqref{e.selfP},\eqref{crn_TFEF}, \eqref{e.selfexpression} with two genes and three EFs. The first gene $\mr X$ refers to both PRRX1 and ZEB1.  The rest of the genome is  modeled by \eqref{e.2EF0}.  Using the above model, we are interested in computing the activation functions for ZEB1 and PRRX1  in an analogous way to \eqref{e.activationfunction}.
	
	Therefore, we first solve the combined competition and self-activation function for one local gene and a global mega-gene to get a function $\Psi(X)$. Then, we write 
	\begin{align*}\Psi\su Z(Z,R)&=\frac{G_{tot}\s Z}{G_{tot}\s Z+G_{tot}\s R} \Psi( (Z/Z_0)^{n\su Z}+(R/R_0)^{n\su R})  ,\\  \Psi\su R(Z,R)&=\frac{G_{tot}\s Z}{G_{tot}\s Z+G_{tot}\s R}  \Psi( (Z/Z_0)^{n\su Z}+(R/R_0)^{n\su R}) ,\end{align*}
	for some $R_0,Z_0>0$, and integers $n_Z,n_R\ge 1$.

	\paragraph{Local GRN} In this case, the GRN is identical to the one presented in the previous subsection except for the new self-activation functions where $\Psi_Z H\su{Z\to Z}(Z)$,$\Psi_R H\su{R\to R}(R)$ are replaced by $\Psi_Z(Z,R),\Psi\su{R}(Z,R)$.  %
	Using \eqref{ss.X},\eqref{ss.X2},\eqref{ss.mu}, we can write the steady state equations as follows:
	\begin{align*}
		0&=k\su{+Z} \Psi\su Z(Z,R)  H \su{S\to Z}(S) \frac{L\su{Z}\s{\mu}(\mu)}{Y\su{Z}\s m(\mu)} - k\su{-Z} Z, \\
		0&=k\su{+R} \Psi\su R(Z,R)  H\su{S\dashv R}(S) H\su{T\to R}(T)- {k\su{-R}} R , \\
		0 &= I\su{\mathrm{ext}} +k\su{+S}  H\su{R\dashv S}(R) H\su{T\to S}(T) - k\su{-S} S, \\
		0&=  k\su{+T}  \frac{L\su{T}\s{\mu}(\mu)}{Y\su{T}\s m(\mu)}  - {k\su{-T}} T, \\
		0 & =  k\su{+\mu}  H\su{Z\dashv\mu}(Z)  H \su{S\dashv \mu}(S)      - k\su{-\mu} \mu -   \frac{Y\su{Z}\s\mu(\mu)\Psi\su Z   H\su{Z\to Z}(Z)  H \su{S\to Z}(S)}{Y\su{Z}\s m(\mu)}  -  \frac{Y\su{T}\s\mu(\mu)}{Y\su{T}\s m(\mu)}.  
	\end{align*}

	\paragraph{Parameters} Here we report the parameters used to generate the simulation in Figure 5-b. All the parameters are identical to the ones reported in the previous subsection. We only need to report the additional parameters that characterize the interaction between the epigenetic competition circuit and self-activation loop. Those parameters are listed below.
	\[
	\begin{array}{rl}\mr T+\mr G_{0X}\s1 \hspace{-0.1in} & \xrightleftharpoons[425.8]{115.6}  \mr G_{TX}\s1 \xrightarrow{127.34}  \mr G_{+X}\s1+ \mr  T, ~ \mr G_{+X}\s1  \xrightarrow{407.97} \mr G_{0X}\s1 \\ \mr  T'+\mr G_{0X}\s1 \hspace{-0.1in} & \xrightleftharpoons[93.91]{729.53} \mr G_{T'X}\s1 \xrightarrow{407.97}  \mr G_{+X}\s1+\mr  T',
	\end{array}\]
	The reactions describing the binding/unbinding of $\mr X$ are listed below:
	\[ \begin{array}{rl}
		\mr X+ \mr G_{00}\s 1 & \xrightleftharpoons[1]{1} \mr G_{0X}\s 1, ~ \mr X+ \mr G_{T0}\s 1   \xrightleftharpoons[1]{1} \mr G_{TX}\s 1,  \\
		\mr X+ \mr G_{T'0}\s 1  & \xrightleftharpoons[1]{1} \mr G_{T'X}\s 1, ~  \mr X+ \mr G_{+0}\s 1   \xrightleftharpoons[1]{1} \mr G_{+X}\s 1.
	\end{array}  \]
	Finally, we have $\rho_X=1$.

	{  \paragraph{Local stability of the steady states} Since the experimental results are  reported at the steady state, the simulation results of our model are also reported   at the steady state. Our model does not include assumptions on the relative time-scale separation between the state variables since it has no impact on the existence of steady states. However, for a general nonlinear system, it is theoretically possible for the asymptotic stability of a steady state to be lost when its subsystems evolve on different time-scales.   To preclude this possibility, we have performed additional computations to verify that the reported  steady  states are asymptotically stable when the GRN states and the EFs evolve on different time-scales, and also with different time-scales for the EFs (which are responsible for writing/erasing of the histone marks).  This is motivated by the observation that histone modification marks can have different half-lives depending on the type of modification \cite{barth10}.
		
		More concretely, let us write the overall dynamical model as follows:
		\begin{equation}\label{dynsys}
			\begin{array}{rl}		\dot x&=f(x,y), \\
				\dot y&=\mathcal{E} g(x,y) \end{array}
		\end{equation}
		where $x=[Z,\mu,R,S,T]^T$, $y=[P,T,T']$, and $\mathcal E$ is defined as follows:
		\[\mathcal E = \varepsilon_0 \begin{bmatrix} \varepsilon_P  & 0 &0 \\ 0 & \varepsilon_T & 0 \\ 0 & 0 & \varepsilon_{T'}\end{bmatrix},\]
		where $\varepsilon_0,\varepsilon_P, \varepsilon_T, \varepsilon_{T'}>0$. The parameter $\varepsilon_0$ controls the relative time-scale separation between the GRN state variables and the EFs. The other three parameters $\varepsilon_P, \varepsilon_T, \varepsilon_{T'}$ control the relative time-scale separation between the different EFs.   For instance, a small $\varepsilon_P$ and large $\varepsilon_{T},\varepsilon_{T'}$ mean that the kinetics of $P$ are slow compared to $T,T'$.
		
		To test stability, we evaluate   the Jacobian of \eqref{dynsys} at the steady state of interest, and check that it is Hurwitz, i.e., we check that all the eigenvalues have negative real parts.  
		
		We performed this calculation for the five knockout scenarios: control, PRC2-KO, KMT2D-KO, $T'$-KO, and (PRC2,KMT2D)-KO. Then, we have calibrated $\varepsilon_0$ so that the two subsystems in \eqref{dynsys} are in the same time-scale judged by having the corresponding eigenvalues  in the same order of magnitude (for the control case). Afterwards, we test stability for eight different cases with $\varepsilon_P,\varepsilon_T,\varepsilon_{T'}$ being either high or low. More precisely, we let  $(\varepsilon_P,\varepsilon_T,\varepsilon_{T'}) \in \{0.01,1\}^3$. In all the tested cases and for all the steady states, the Jacobian has been verified to be Hurwitz. In other words, local stability of the steady-states is verified across widely different time scale.

	}


\begin{thebibliography}{10}
 	
 	\bibitem{ama_book}
 	Wolpert L, Tickle C, Arias AM (2019) {\em Principles of development}.
 	\newblock (Oxford University Press), Sixth edition.
 	
 	\bibitem{emt_2016}
 	Nieto MA, Huang RYJ, Jackson RA, Thiery JP (2016) {EMT}: 2016.
 	\newblock {\em Cell} 166(1):21--45.
 	
 	\bibitem{Wu2012}
 	Wu CY, Tsai YP, Wu MZ, Teng SC, Wu KJ (2012) Epigenetic reprogramming and
 	post-transcriptional regulation during the epithelial{\textendash}mesenchymal
 	transition.
 	\newblock {\em Trends in Genetics} 28(9):454--463.
 	
 	\bibitem{scheel}
 	Eichelberger L, et~al. (2020) Maintenance of epithelial traits and resistance
 	to mesenchymal reprogramming promote proliferation in metastatic breast
 	cancer.
 	\newblock {\em BioRxiv}.
 	
 	\bibitem{chaffer}
 	Chaffer CL, et~al. (2013) Poised chromatin at the {ZEB}1 promoter enables
 	breast cancer cell plasticity and enhances tumorigenicity.
 	\newblock {\em Cell} 154(1):61--74.
 	
 	\bibitem{yun22}
 	Zhang Y, et~al. (2022) Genome-wide crispr screen identifies prc2 and
 	kmt2d-compass as regulators of distinct emt trajectories that contribute
 	differentially to metastasis.
 	\newblock {\em Nature Cell Biology} p. 1–11.
 	
 	\bibitem{karlebach08}
 	Karlebach G, Shamir R (2008) Modelling and analysis of gene regulatory
 	networks.
 	\newblock {\em Nature reviews Molecular cell biology} 9(10):770--780.
 	
 	\bibitem{huang05}
 	Huang S, Eichler G, Bar-Yam Y, Ingber DE (2005) Cell fates as high-dimensional
 	attractor states of a complex gene regulatory network.
 	\newblock {\em Physical Review Letters} 94(12):128701.
 	
 	\bibitem{racipe}
 	Huang B, et~al. (2017) Interrogating the topological robustness of gene
 	regulatory circuits by randomization.
 	\newblock {\em PLOS Computational Biology} 13(3):e1005456.
 	
 	\bibitem{kauffman71}
 	Kauffman S (1971) Gene regulation networks: A theory for their global structure
 	and behaviors in {\em Current topics in developmental biology}.
 	\newblock (Elsevier) Vol.{}~6, pp. 145--182.
 	
 	\bibitem{alon}
 	Alon U (2006) {\em An Introduction to Systems Biology}.
 	\newblock (CRC press).
 	
 	\bibitem{lu13b}
 	Lu M, et~al. (2013) Tristability in cancer-associated micro{RNA-TF} chimera
 	toggle switch.
 	\newblock {\em The Journal of Physical Chemistry B} 117(42):13164--13174.
 	
 	\bibitem{olariu16}
 	Olariu V, L{\"o}vkvist C, Sneppen K (2016) Nanog, {Oct4} and {Tet1} interplay
 	in establishing pluripotency.
 	\newblock {\em Scientific reports} 6(1):1--11.
 	
 	\bibitem{chen21}
 	Chen T, Ali Al-Radhawi M, Sontag ED (2021) A mathematical model exhibiting the
 	effect of {DNA} methylation on the stability boundary in cell-fate networks.
 	\newblock {\em Epigenetics} 16(4):436--457.
 	
 	\bibitem{thalheim17}
 	Thalheim T, Herberg M, Loeffler M, Galle J (2017) The regulatory capacity of
 	bivalent genes—a theoretical approach.
 	\newblock {\em International Journal of Molecular Sciences} 18(5):1069.
 	
 	\bibitem{Zhang19}
 	Zhang Y, Liu N, Lin W, Li C (2019) Quantifying the interplay between genetic
 	and epigenetic regulations in stem cell development.
 	\newblock {\em New Journal of Physics} 21(10):103042.
 	
 	\bibitem{Alarcon21}
 	Alarcón T, Sardanyés J, Guillamon A, Menendez JA (2021) Bivalent chromatin as
 	a therapeutic target in cancer: An in silico predictive approach for
 	combining epigenetic drugs.
 	\newblock {\em PLOS Computational Biology} 17(6):e1008408.
 	
 	\bibitem{bruno22}
 	Bruno S, Williams RJ, Del~Vecchio D (2022) Epigenetic cell memory: The gene’s
 	inner chromatin modification circuit.
 	\newblock {\em PLoS computational biology} 18(4):e1009961.
 	
 	\bibitem{sneppen2019}
 	Sneppen K, Ringrose L (2019) Theoretical analysis of {Polycomb}-trithorax
 	systems predicts that poised chromatin is bistable and not bivalent.
 	\newblock {\em Nature Communications} 10(1):2133.
 	
 	\bibitem{zhao21}
 	Zhao W, Qiao L, Yan S, Nie Q, Zhang L (2021) Mathematical modeling of histone
 	modifications reveals the formation mechanism and function of bivalent
 	chromatin.
 	\newblock {\em IScience} 24(7):102732.
 	
 	\bibitem{ringrose20}
 	Reinig J, Ruge F, Howard M, Ringrose L (2020) A theoretical model of
 	polycomb/trithorax action unites stable epigenetic memory and dynamic
 	regulation.
 	\newblock {\em Nature communications} 11(1):1--16.
 	
 	\bibitem{systemsbiologyEMT}
 	Jolly MK, Levine H (2017) Computational systems biology of
 	epithelial-hybrid-mesenchymal transitions.
 	\newblock {\em Current Opinion in Systems Biology} 3:1--6.
 	
 	\bibitem{Jia2019}
 	Jia W, Deshmukh A, Mani SA, Jolly MK, Levine H (2019) A possible role for
 	epigenetic feedback regulation in the dynamics of the
 	epithelial{\textendash}mesenchymal transition ({EMT}).
 	\newblock {\em Physical Biology} 16(6):066004.
 	
 	\bibitem{Jia2020}
 	Jia W, et~al. (2020) Epigenetic feedback and stochastic partitioning during
 	cell division can drive resistance to {EMT}.
 	\newblock {\em Oncotarget} 11(27):2611--2624.
 	
 	\bibitem{karacosta19}
 	Karacosta LG, et~al. (2019) Mapping lung cancer epithelial-mesenchymal
 	transition states and trajectories with single-cell resolution.
 	\newblock {\em Nature communications} 10(1):1--15.
 	
 	\bibitem{Piunti16}
 	Piunti A, Shilatifard A (2016) Epigenetic balance of gene expression by
 	{Polycomb} and {COMPASS} families.
 	\newblock {\em Science} 352(6290):aad9780.
 	
 	\bibitem{schuettengruber17}
 	Schuettengruber B, Bourbon HM, Di~Croce L, Cavalli G (2017) Genome regulation
 	by polycomb and trithorax: 70 years and counting.
 	\newblock {\em Cell} 171(1):34--57.
 	
 	\bibitem{Margueron11}
 	Margueron R, Reinberg D (2011) The {Polycomb} complex {PRC}2 and its mark in
 	life.
 	\newblock {\em Nature} 469(7330):343–349.
 	
 	\bibitem{conway15}
 	Conway E, Healy E, Bracken AP (2015) {PRC}2 mediated {H3K27} methylations in
 	cellular identity and cancer.
 	\newblock {\em Current opinion in cell biology} 37:42--48.
 	
 	\bibitem{bracken06}
 	Bracken AP, Dietrich N, Pasini D, Hansen KH, Helin K (2006) Genome-wide mapping
 	of {Polycomb} target genes unravels their roles in cell fate transitions.
 	\newblock {\em Genes \& development} 20(9):1123--1136.
 	
 	\bibitem{mohn08}
 	Mohn F, et~al. (2008) Lineage-specific polycomb targets and de novo {DNA}
 	methylation define restriction and potential of neuronal progenitors.
 	\newblock {\em Molecular cell} 30(6):755--766.
 	
 	\bibitem{Shilatifard12}
 	Shilatifard A (2012) The {COMPASS} family of histone {H3K4} methylases:
 	Mechanisms of regulation in development and disease pathogenesis.
 	\newblock {\em Annual Review of Biochemistry} 81(1):65–95.
 	
 	\bibitem{wu08}
 	Wu M, et~al. (2008) Molecular regulation of {H3K4} trimethylation by {Wdr82}, a
 	component of human {S}et1/{COMPASS}.
 	\newblock {\em Molecular and cellular biology} 28(24):7337--7344.
 	
 	\bibitem{hu13}
 	Hu D, et~al. (2013) The {MLL}3/{MLL}4 branches of the {COMPASS} family function
 	as major histone {H3K4} monomethylases at enhancers.
 	\newblock {\em Molecular and cellular biology} 33(23):4745--4754.
 	
 	\bibitem{froimchuk17}
 	Froimchuk E, Jang Y, Ge K (2017) {Histone {H3} lysine 4 methyltransferase
 		{KMT2D}}.
 	\newblock {\em Gene} 627:337--342.
 	
 	\bibitem{hu17}
 	Hu D, et~al. (2017) Not all {H3K4} methylations are created equal:
 	Mll2/{COMPASS} dependency in primordial germ cell specification.
 	\newblock {\em Molecular cell} 65(3):460--475.
 	
 	\bibitem{zhang15}
 	Zhang J, et~al. (2015) Disruption of {KMT2D} perturbs germinal center b cell
 	development and promotes lymphomagenesis.
 	\newblock {\em Nature Medicine} 21(10):1190–1198.
 	
 	\bibitem{dhar18}
 	Dhar SS, et~al. (2018) {MLL}4 is required to maintain broad {H3K4}me3 peaks and
 	super-enhancers at tumor suppressor genes.
 	\newblock {\em Molecular Cell} 70(5):825--841.e6.
 	
 	\bibitem{ang16}
 	Ang SY, et~al. (2016) {KMT2D} regulates specific programs in heart development
 	via histone {H3} lysine 4 di-methylation.
 	\newblock {\em Development} 143(5):810–821.
 	
 	\bibitem{wang16}
 	Wang C, et~al. (2016) Enhancer priming by {H3K4} methyltransferase {MLL}4
 	controls cell fate transition.
 	\newblock {\em Proceedings of the National Academy of Sciences}
 	113(42):11871–11876.
 	
 	\bibitem{Douillet20}
 	Douillet D, et~al. (2020) Uncoupling histone {H3K4} trimethylation from
 	developmental gene expression via an equilibrium of {COMPASS}, {Polycomb} and
 	{DNA} methylation.
 	\newblock {\em Nature Genetics} 52(6):615–625.
 	
 	\bibitem{steffen14}
 	Steffen PA, Ringrose L (2014) What are memories made of? how {Polycomb} and
 	trithorax proteins mediate epigenetic memory.
 	\newblock {\em Nature Reviews Molecular Cell Biology} 15(5):340--356.
 	
 	\bibitem{klymenko04}
 	Klymenko T, M{\"u}ller J (2004) The histone methyltransferases {Trithorax} and
 	{Ash1} prevent transcriptional silencing by {Polycomb} group proteins.
 	\newblock {\em EMBO reports} 5(4):373--377.
 	
 	\bibitem{tie09}
 	Tie F, et~al. (2009) {CBP}-mediated acetylation of histone {H3} lysine 27
 	antagonizes drosophila {Polycomb} silencing.
 	\newblock {\em Development} 136(18):3131–3141.
 	
 	\bibitem{schmitges11}
 	Schmitges FW, et~al. (2011) Histone methylation by {PRC}2 is inhibited by
 	active chromatin marks.
 	\newblock {\em Molecular cell} 42(3):330--341.
 	
 	\bibitem{pasini10}
 	Pasini D, et~al. (2010) Characterization of an antagonistic switch between
 	histone {H3} lysine 27 methylation and acetylation in the transcriptional
 	regulation of {Polycomb} group target genes.
 	\newblock {\em Nucleic Acids Research} 38(15):4958–4969.
 	
 	\bibitem{banerjee16}
 	Tie F, et~al. (2016) {Polycomb} inhibits histone acetylation by {CBP} by
 	binding directly to its catalytic domain.
 	\newblock {\em Proceedings of the National Academy of Sciences}
 	113(6):E744–E753.
 	
 	\bibitem{blanco20}
 	Blanco E, Gonz{\'a}lez-Ram{\'\i}rez M, Alcaine-Colet A, Aranda S, Di~Croce L
 	(2020) The bivalent genome: characterization, structure, and regulation.
 	\newblock {\em Trends in Genetics} 36(2):118--131.
 	
 	\bibitem{shema16}
 	Shema E, et~al. (2016) Single-molecule decoding of combinatorially modified
 	nucleosomes.
 	\newblock {\em Science} 352(6286):717--721.
 	
 	\bibitem{pasini08}
 	Pasini D, et~al. (2008) Coordinated regulation of transcriptional repression by
 	the rbp2 h3k4 demethylase and polycomb-repressive complex 2.
 	\newblock {\em Genes \& development} 22(10):1345--1355.
 	
 	\bibitem{jin17}
 	Jin Y, et~al. (2017) Lsd1 collaborates with ezh2 to regulate expression of
 	interferon-stimulated genes.
 	\newblock {\em Biomedicine \& Pharmacotherapy} 88:728--737.
 	
 	\bibitem{guo21}
 	Guo Y, Zhao S, Wang GG (2021) Polycomb gene silencing mechanisms: {PRC2}
 	chromatin targeting, {H3K27me3} “readout”, and phase separation-based
 	compaction.
 	\newblock {\em Trends in Genetics} 37(6):547–565.
 	
 	\bibitem{Holoch17}
 	Holoch D, Margueron R (2017) Mechanisms regulating {PRC}2 recruitment and
 	enzymatic activity.
 	\newblock {\em Trends in Biochemical Sciences} 42(7):531–542.
 	
 	\bibitem{wiles17}
 	Wiles ET, Selker EU (2017) {H3K27} methylation: a promiscuous repressive
 	chromatin mark.
 	\newblock {\em Current Opinion in Genetics \& Development} 43:31–37.
 	
 	\bibitem{davidovich21}
 	Uckelmann M, Davidovich C (2021) Not just a writer: {PRC}2 as a chromatin
 	reader.
 	\newblock {\em Biochemical Society Transactions} 49(3):1159–1170.
 	
 	\bibitem{beltran16}
 	Beltran M, et~al. (2016) The interaction of {PRC2} with {RNA} or chromatin is
 	mutually antagonistic.
 	\newblock {\em Genome research} 26(7):896--907.
 	
 	\bibitem{yuan12}
 	Yuan W, et~al. (2012) Dense chromatin activates polycomb repressive complex 2
 	to regulate {H3} lysine 27 methylation.
 	\newblock {\em science} 337(6097):971--975.
 	
 	\bibitem{perez11}
 	P{\'e}rez L, et~al. (2011) Enhancer-pre communication contributes to the
 	expansion of gene expression domains in proliferating primordia.
 	\newblock {\em Development} 138(15):3125--3134.
 	
 	\bibitem{gaydos12}
 	Gaydos L, Rechtsteiner A, Egelhofer T, Carroll C, Strome S (2012) Antagonism
 	between {MES}-4 and polycomb repressive complex 2 promotes appropriate gene
 	expression in {C.} elegans germ cells.
 	\newblock {\em Cell Reports} 2(5):1169–1177.
 	
 	\bibitem{lu16}
 	Lu C, et~al. (2016) Histone {H3K36} mutations promote sarcomagenesis through
 	altered histone methylation landscape.
 	\newblock {\em Science} 352(6287):844–849.
 	
 	\bibitem{chaouch21}
 	Chaouch A, et~al. (2021) {Histone H3.3 K27M and K36M mutations de-repress
 		transposable elements through perturbation of antagonistic chromatin marks}.
 	\newblock {\em Molecular Cell} 81(23):4876--4890.e7.
 	
 	\bibitem{weber21}
 	Weber CM, et~al. (2021) {mSWI/SNF promotes {Polycomb} repression both directly
 		and through genome-wide redistribution}.
 	\newblock {\em Nature Structural \& Molecular Biology} 28(6):501–511.
 	
 	\bibitem{mortimer19}
 	Mortimer T, et~al. (2019) Redistribution of {EZH} 2 promotes malignant
 	phenotypes by rewiring developmental programmes.
 	\newblock {\em EMBO reports} 20(10):e48155.
 	
 	\bibitem{khazaei20}
 	Khazaei S, et~al. (2020) {H3.3 G34W} promotes growth and impedes
 	differentiation of osteoblast-like mesenchymal progenitors in giant cell
 	tumor of bone.
 	\newblock {\em Cancer Discovery} 10(12):1968–1987.
 	
 	\bibitem{hanna22}
 	Hanna CW, et~al. (2022) Loss of histone methyltransferase {SETD1B} in oogenesis
 	results in the redistribution of genomic histone 3 lysine 4 trimethylation.
 	\newblock {\em Nucleic Acids Research} 50(4):1993–2004.
 	
 	\bibitem{walter16}
 	Walter M, Teissandier A, Pérez-Palacios R, Bourc’his D (2016) An epigenetic
 	switch ensures transposon repression upon dynamic loss of {DNA} methylation
 	in embryonic stem cells.
 	\newblock {\em eLife} 5:e11418.
 	
 	\bibitem{lu13}
 	Lu M, Jolly MK, Levine H, Onuchic JN, Ben-Jacob E (2013) {MicroRNA}-based
 	regulation of epithelial-hybrid-mesenchymal fate determination.
 	\newblock {\em Proceedings of the National Academy of Sciences}
 	110(45):18144–18149.
 	
 	\bibitem{fazilaty19}
 	Fazilaty H, et~al. (2019) A gene regulatory network to control {EMT} programs
 	in development and disease.
 	\newblock {\em Nature Communications} 10(1):1--16.
 	
 	\bibitem{bocci19}
 	Bocci F, Levine H, Onuchic JN, Jolly MK (2019) Deciphering the dynamics of
 	epithelial-mesenchymal transition and cancer stem cells in tumor progression.
 	\newblock {\em Current Stem Cell Reports} 5(1):11–21.
 	
 	\bibitem{fagan19}
 	Fagan RJ, Dingwall AK (2019) {COMPASS} ascending: Emerging clues regarding the
 	roles of {MLL}3/{KMT2C} and {MLL}2/{KMT2D} proteins in cancer.
 	\newblock {\em Cancer letters} 458:56--65.
 	
 	\bibitem{kim16}
 	Kim KH, Roberts CW (2016) Targeting {EZH}2 in cancer.
 	\newblock {\em Nature medicine} 22(2):128--134.
 	
 	\bibitem{duan20}
 	Duan R, Du W, Guo W (2020) {EZH}2: a novel target for cancer treatment.
 	\newblock {\em Journal of hematology \& oncology} 13(1):1--12.
 	
 	\bibitem{eich20}
 	Eich ML, Athar M, Ferguson JE, Varambally S (2020) {EZH}2-targeted therapies in
 	cancer: hype or a reality.
 	\newblock {\em Cancer research} 80(24):5449--5458.
 	
 	\bibitem{deleris12}
 	Deleris A, et~al. (2012) Loss of the {DNA} methyltransferase {MET1} induces
 	{H3K9} hypermethylation at {PcG} target genes and redistribution of {H3K27}
 	trimethylation to transposons in {A}rabidopsis thaliana.
 	\newblock {\em PLOS Genetics} 8(11):e1003062.
 	
 	\bibitem{basenko15}
 	Basenko EY, et~al. (2015) Genome-wide redistribution of {{H3K27me3}} is linked
 	to genotoxic stress and defective growth.
 	\newblock {\em Proceedings of the National Academy of Sciences}
 	112(46):E6339–E6348.
 	
 	\bibitem{reddington13}
 	Reddington JP, et~al. (2013) Redistribution of {H3K27me3} upon {{H3K27me3}}
 	hypomethylation results in de-repression of {Polycomb} target genes.
 	\newblock {\em Genome Biology} 14(3):R25.
 	
 	\bibitem{jamieson16}
 	Jamieson K, et~al. (2016) Loss of {HP}1 causes depletion of {H3K27me3} from
 	facultative heterochromatin and gain of {H3K27me2} at constitutive
 	heterochromatin.
 	\newblock {\em Genome Research} 26(1):97–107.
 	
 	\bibitem{ren20}
 	Ren W, et~al. (2020) Disruption of {ATRX-RNA} interactions uncovers roles in
 	{ATRX} localization and {PRC}2 function.
 	\newblock {\em Nature Communications} 11(1):2219.
 	
 	\bibitem{morgan15}
 	Morgan MA, Shilatifard A (2015) Chromatin signatures of cancer.
 	\newblock {\em Genes \& development} 29(3):238--249.
 	
 	\bibitem{laugesen16}
 	Laugesen A, H{\o}jfeldt JW, Helin K (2016) Role of the polycomb repressive
 	complex 2 ({PRC}2) in transcriptional regulation and cancer.
 	\newblock {\em Cold Spring Harbor perspectives in medicine} 6(9):a026575.
 	
 	\bibitem{dockerill21}
 	Dockerill M, Gregson C, O’Donovan DH (2021) Targeting {PRC}2 for the
 	treatment of cancer: an updated patent review (2016-2020).
 	\newblock {\em Expert Opinion on Therapeutic Patents} 31(2):119--135.
 	
 	\bibitem{sze16}
 	Sze CC, Shilatifard A (2016) {MLL}3/{MLL}4/{COMPASS} family on epigenetic
 	regulation of enhancer function and cancer.
 	\newblock {\em Cold Spring Harbor perspectives in medicine} 6(11):a026427.
 	
 	\bibitem{dawkins16}
 	Dawkins JB, et~al. (2016) Reduced expression of histone methyltransferases
 	{KMT2C} and {KMT2D} correlates with improved outcome in pancreatic ductal
 	adenocarcinoma.
 	\newblock {\em Cancer research} 76(16):4861--4871.
 	
 	\bibitem{gala18}
 	Gala K, et~al. (2018) {KMT2C} mediates the estrogen dependence of breast cancer
 	through regulation of er$\alpha$ enhancer function.
 	\newblock {\em Oncogene} 37(34):4692--4710.
 	
 	\bibitem{xiong19}
 	Xiong W, et~al. (2019) {MLL}3 enhances the transcription of {PD-L1} and
 	regulates anti-tumor immunity.
 	\newblock {\em Biochimica et Biophysica Acta (BBA)-Molecular Basis of Disease}
 	1865(2):454--463.
 	
 	\bibitem{yu20}
 	Yu Q, et~al. (2020) Small molecule inhibitors of the prostate cancer target
 	{KMT2D}.
 	\newblock {\em Biochemical and Biophysical Research Communications}
 	533(3):540--547.
 	
 	\bibitem{dhar21}
 	Dhar SS, Lee MG (2021) Cancer-epigenetic function of the histone
 	methyltransferase {KMT2D} and therapeutic opportunities for the treatment of
 	{KMT2D}-deficient tumors.
 	\newblock {\em Oncotarget} 12(13):1296.
 	
 	\bibitem{mani17}
 	Taube JH, et~al. (2017) The {H3K27me3}-demethylase {KDM6A} is suppressed in
 	breast cancer stem-like cells, and enables the resolution of bivalency during
 	the mesenchymal-epithelial transition.
 	\newblock {\em Oncotarget} 8(39):65548–65565.
 	
 	\bibitem{qian17}
 	Qian Y, Huang HH, Jim{\'e}nez JI, Del~Vecchio D (2017) Resource competition
 	shapes the response of genetic circuits.
 	\newblock {\em ACS Synthetic Biology} 6(7):1263--1272.
 	
 	\bibitem{MA_LCSS22}
 	Al-Radhawi MA, Del~Vecchio D, Sontag ED (2022) Identifying competition
 	phenotypes in synthetic biochemical circuits.
 	\newblock {\em bioRxiv}.
 	
 	\bibitem{shuyi18}
 	Zhang S, Voigt CA (2018) Engineered {dCas9} with reduced toxicity in bacteria:
 	implications for genetic circuit design.
 	\newblock {\em Nucleic Acids Research} 46(20):11115--11125.
 	
 	\bibitem{raveh16}
 	Raveh A, Margaliot M, Sontag ED, Tuller T (2016) A model for competition for
 	ribosomes in the cell.
 	\newblock {\em Journal of The Royal Society Interface} 13(116):20151062.
 	
 	\bibitem{miller21}
 	Miller J, {Ali Al-Radhawi} M, Sontag ED (2021) Mediating ribosomal competition
 	by splitting pools.
 	\newblock {\em IEEE Control Systems Letters} 5(5):1555--1560.
 	
 	\bibitem{bosia}
 	Bosia C, et~al. (2017) {RNAs} competing for {microRNAs} mutually influence
 	their fluctuations in a highly non-linear {microRNA}-dependent manner in
 	single cells.
 	\newblock {\em Genome Biology} 18(1).
 	
 	\bibitem{salmena}
 	Salmena L, Poliseno L, Tay Y, Kats L, Pandolfi PP (2011) A {ceRNA} hypothesis:
 	The rosetta stone of a hidden {RNA} language?
 	\newblock {\em Cell} 146(3):353--358.
 	
 	\bibitem{gsea}
 	Subramanian A, et~al. (2005) Gene set enrichment analysis: A knowledge-based
 	approach for interpreting genome-wide expression profiles.
 	\newblock {\em Proceedings of the National Academy of Sciences}
 	102(43):15545--15550.
 	
 	\bibitem{ontology}
 	Mi H, Muruganujan A, Ebert D, Huang X, Thomas PD (2018) {PANTHER} version 14:
 	more genomes, a new {PANTHER} {GO}-slim and improvements in enrichment
 	analysis tools.
 	\newblock {\em Nucleic Acids Research} 47(D1):D419--D426.
 	
 	\bibitem{shanprc2}
 	Shan Y, et~al. (2017) Prc2 specifies ectoderm lineages and maintains
 	pluripotency in primed but not na{\"\i}ve escs.
 	\newblock {\em Nature communications} 8(1):1--14.
 	
 \end{thebibliography}

{\footnotesize
	\begin{thebibliography}{10}
		\providecommand{\url}[1]{\texttt{#1}}
		\providecommand{\urlprefix}{URL }
		\expandafter\ifx\csname urlstyle\endcsname\relax
		\providecommand{\doi}[1]{doi:\discretionary{}{}{}#1}\else
		\providecommand{\doi}{doi:\discretionary{}{}{}\begingroup
			\urlstyle{rm}\Url}\fi
		\providecommand{\eprint}[2][]{\url{#2}}
		
		\bibitem{feinberg87}
		M.~Feinberg.
		\newblock Chemical reaction network structure and the stability of complex
		isothermal reactors--{I}. {T}he deficiency zero and deficiency one theorems.
		\newblock \emph{Chemical Engineering Science}, volume~42(10):pages 2229--2268,
		1987.
		
		\bibitem{erdi89}
		P.~{\'E}rdi and J.~T{\'o}th.
		\newblock \emph{Mathematical models of chemical reactions: theory and
			applications of deterministic and stochastic models}.
		\newblock Manchester University Press, 1989.
		
		\bibitem{MA_LEARN}
		M.~Ali Al-Radhawi, D.~Angeli, and E.~D. Sontag.
		\newblock A computational framework for a {L}yapunov-enabled analysis of
		biochemical reaction networks.
		\newblock \emph{PLoS Computational Biology}, volume~16(2):page e1007681, 2020.
		
		\bibitem{craciun05}
		G.~Craciun and M.~Feinberg.
		\newblock Multiple equilibria in complex chemical reaction networks: I. the
		injectivity property.
		\newblock \emph{SIAM Journal on Applied Mathematics}, pages 1526--1546, 2005.
		
		\bibitem{banaji07}
		M.~Banaji, P.~Donnell, and S.~Baigent.
		\newblock P matrix properties, injectivity, and stability in chemical reaction
		systems.
		\newblock \emph{SIAM Journal on Applied Mathematics}, volume~67(6):pages
		1523--1547, 2007.
		
		\bibitem{lu13b}
		M.~Lu, M.~K. Jolly, R.~Gomoto, B.~Huang, et~al.
		\newblock Tristability in cancer-associated micro{RNA-TF} chimera toggle
		switch.
		\newblock \emph{The Journal of Physical Chemistry B}, volume 117(42):pages
		13164--13174, 2013.
		
		\bibitem{racipe}
		B.~Huang, M.~Lu, D.~Jia, E.~Ben-Jacob, et~al.
		\newblock Interrogating the topological robustness of gene regulatory circuits
		by randomization.
		\newblock \emph{PLOS Computational Biology}, volume~13(3):page e1005456, 2017.
		\newblock \doi{10.1371/journal.pcbi.1005456}.
		
		\bibitem{lu13}
		M.~Lu, M.~K. Jolly, H.~Levine, J.~N. Onuchic, et~al.
		\newblock {MicroRNA}-based regulation of epithelial-hybrid-mesenchymal fate
		determination.
		\newblock \emph{Proceedings of the National Academy of Sciences}, volume
		110(45):page 18144–18149, 2013.
		\newblock \doi{10.1073/pnas.1318192110}.
		
		\bibitem{MA_springer22}
		M.~A. Al-Radhawi and E.~D. Sontag.
		\newblock Analysis of a reduced model of epithelial–mesenchymal fate
		determination in cancer metastasis as a singularly-perturbed monotone system.
		\newblock In C.~Beattie, P.~Benner, M.~Embree, S.~Gugercin, et~al., editors,
		\emph{Realization and Model Reduction of Dynamical Systems}. Springer-Verlag,
		2022.
		
		\bibitem{barth10}
		T.~K. Barth and A.~Imhof.
		\newblock Fast signals and slow marks: the dynamics of histone modifications.
		\newblock \emph{Trends in biochemical sciences}, volume~35(11):pages 618--626,
		2010.
		
	\end{thebibliography}
}
\end{document}